\documentclass[a4paper,iop,twocolumn,twocolappendix,numberedappendix]{emulateapj} 
\usepackage[breaklinks,colorlinks=true,linkcolor=blue,anchorcolor=blue,citecolor=blue,urlcolor=blue,draft=false]{hyperref}
\usepackage{color}
\usepackage{graphicx}
\usepackage{amsmath} 
\usepackage{amssymb}
\usepackage{mathrsfs}
\usepackage{placeins}
\usepackage{times}
\usepackage{xcolor}
\usepackage{xspace}
\shorttitle{The Dynamical State of Abell 370}
\shortauthors{Molnar, Ueda \& Umetsu}

\newcommand{\simless} 
     {\ensuremath{\lower 3pt\hbox{$\rlap{\raise5pt\hbox{$\char'074$}}\mathchar"7218$}}}
\newcommand{\simgreat}
     {\ensuremath{\lower 3pt\hbox{$\rlap{\raise5pt\hbox{$\char'076$}}\mathchar"7218$}}}

\newcommand{\simgt}{\lower.5ex\hbox{$\; \buildrel > \over \sim \;$}}
\newcommand{\simlt}{\lower.5ex\hbox{$\; \buildrel < \over \sim \;$}}

\newcommand{\etal}{{\it et al.}}

\newcommand{\LCDM}{{{\sc $\Lambda$CDM}}\xspace}

\newcommand{\HE}{hydrostatic equilibrium}

\newcommand{\ASEC}{\ensuremath{\arcsec}\xspace}
\newcommand{\AMIN}{\ensuremath{\arcmin}\xspace}

\newcommand{\KMSEC}{km\,s$^{-1}$\xspace}

\newcommand{\Mvir}{{\ensuremath{M_\mathrm{vir}}}}
\newcommand{\Cvir}{ {\ensuremath{c_\mathrm{vir}}} }

\newcommand{\percent}{\ensuremath{\%}}

\newcommand*{\ltsim}{\ {\raise-.75ex\hbox{$\buildrel<\over\sim$}}\ }
\newcommand*{\gtsim}{\ {\raise-.75ex\hbox{$\buildrel>\over\simsaus$}}\ }
\newcommand*{\proptosim}{\ {\raise-.75ex\hbox{$\buildrel\propto\over\sim$}}\ }

\newcommand*{\CHANDRA}{\emph{Chandra}\xspace}

\newcommand*{\NBODYHYDRO}{{$N$-body/hydro\-dynamical}\xspace}

\newcommand{\FIGURES}{./}

\newcommand*{\FLASH}{\textsc{flash}\xspace}

\begin{document}

\title{The Dynamical State of the Frontier Fields Galaxy Cluster Abell 370}

\author{Sandor M. Molnar\altaffilmark{1}}
\author{Shutaro Ueda\altaffilmark{1}}
\author{Keiichi Umetsu\altaffilmark{1}}

\altaffiltext{1}{Institute of Astronomy and Astrophysics, Academia Sinica, 
                 No. 1, Section 4, Roosevelt Road, Taipei 10617, Taiwan, R.O.C.}


\begin{abstract}
We study the dynamics of Abell 370 (A370), a highly massive Hubble
Frontier Fields galaxy cluster, using self-consistent three-dimensional
$N$-body/hydrodynamical simulations. 
Our simulations are constrained by X-ray, optical spectroscopic and
 gravitational lensing,  and Sunyaev--Zel'dovich (SZ) effect observations. 
Analyzing archival \CHANDRA observations of A370 and comparing the 
X-ray morphology to the latest gravitational lensing mass reconstruction,
we find offsets of $\sim 30$\,kpc and $\sim 100$\,kpc between the 
two X-ray surface brightness peaks and their nearest mass surface density peaks, 
suggesting that it is a merging system, in agreement with previous studies.
Based on our dedicated binary cluster merger simulations, 
we find that initial conditions of the two progenitors with virial
masses of $1.7 \times 10^{15} \,M_\odot$ and $1.6 \times 10^{15} \,M_\odot$, 
an infall velocity of $3500\,$\KMSEC, and an impact parameter of 100\,kpc
can explain the positions and the offsets between 
the peaks of the X-ray emission and mass surface density, 
the amplitude of the integrated SZ signal, and the observed relative
 line-of-sight velocity.
Moreover, our best model reproduces the observed velocity dispersion of
cluster member galaxies, which supports the large total mass of A370
derived from weak lensing.
Our simulations strongly suggest that A370 is a post major
merger after the second core passage in the infalling phase, just before the third core passage. 
In this phase, the gas has not settled down in the gravitational potential well of the cluster, 
which explains why A370 does not follow closely the galaxy cluster scaling relations.
\end{abstract}

\section{Introduction}
\label{S:Intro}

Abell 370 (A370) is a well-studied galaxy cluster at a redshift of $z = 0.375$
\citep[e.g.,][and references therein]{Tyson1990,Kochanek1990,Miralda-Escude1991,
Umetsu1999,Medezinski2010,SchmidtET2014,TreuET2015,LagattutaET2017,
LotzET2017,LagattutaET2019}.  
It was one of the first galaxy clusters in which gravitational lensing was observed 
\citep{Soucail1987}. The first spectroscopically confirmed giant arc was also discovered in this cluster 
\citep{SoucailET1988}.
A370 was found to be one of the most massive clusters based on weak
gravitational lensing, with a total virial mass of 
$M_\mathrm{vir}\approx 3.3 \times 10^{15}M_\odot$ \citep{Umetsu2011a,Umetsu2011b}.
Because of its large projected mass and high lensing magnification capability of background galaxies,
A370 has been known as a {\em superlens}
characterized by large Einstein radii \citep[e.g., $\theta_\mathrm{Ein}>30\arcsec$ for $z_s=2$;][]{Broadhurst2008},
and selected as one of the six Hubble Frontier
Fields\footnote{https://frontierfields.org} clusters \citep{LotzET2017}.
More recently, A370 has been targeted by the BUFFALO survey
\citep{Steinhardt2020} with {\em Hubble Space Telescope} ({\em HST}),
which will expand the existing area  coverage of the Hubble Frontier
Fields in optical and near-infrared pass bands.

The mass surface density, the galaxy number density,
and X-ray surface brightness show a symmetric bimodal distribution along
the north-south direction  
in the core of A370, suggesting that it is a massive major merger 
with a mass ratio close to unity 
\citep{RichardET2010,LagattutaET2017,StraitET2018ApJ868,DiegoET2018MNRAS473}.
Two brightest cluster galaxies (BCGs) were found in the core of A370.
The BCG located in the south of the core of the cluster is very 
close to the mass peak of the southern cluster component (within a few kpc),
while the northern BCG shows a significant offset
from the northern cluster mass peak ($\sim 50$\,kpc;
e.g., \citealt{RichardET2010,StraitET2018ApJ868}).
\cite{BotteonET2018MNRAS476} used X-ray observations of A370 to search for 
surface brightness discontinuities to identify shocks and contact discontinuities.
They found evidence for surface brightness edges on the western and the eastern 
side of the cluster.
Although the origin of the surface brightness edges is not clear, the
edges might be associated with shocks induced by a merger.

Galaxy cluster scaling relations applied to A370 also point to its dynamical activity.
Based on the total mass $M_{500}$ (all relevant symbols are defined in detail at the end of
this section) estimated from weak lensing \citep{Umetsu2011a},
the X-ray luminosity in the 0.5 -- 2 keV band is a factor of
$\sim2$--$3$ times smaller than that inferred from the
$L_\mathrm{X}$--$M_{500}$  scaling relation of \cite{PrattET2009}.
Moreover, $M_{500}$ estimated by {\em Planck} Sunyaev--Zel'dovich (SZ)
effect observations is a factor of $\sim 2.5$ smaller than the
weak-lensing mass \citep{Umetsu2011a,CoeET2019}. 
All of these pieces of evidence, namely
the discrepancies in mass scaling relations, positional offsets between
the galaxy and mass surface densities, and multiple X-ray brightness
peaks, imply dynamical activity in the cluster.

Motivated by these observational results, here we carry out the first
attempt to model A370  as a binary major merger by performing dedicated
numerical simulations on CPU clusters using our three-dimensional (3D)
\NBODYHYDRO code based on \FLASH
\citep{MolnarET2012ApJ748,Molnar2013ApJ774,Molnar2013ApJ779,MolnarBroadhurst2015,MolnarBroadhurst2017,MolnarBroadhurst2018} 
developed at the University of Chicago \citep{Fryxell2000ApJS131p273}.

The structure of this paper is as follows.
In Section~\ref{S:Observations}, we summarize the results 
from multi-wavelength observations of A370 including our results
based on re-analyzing existing \CHANDRA X-ray observations.
We describe our simulation setup to obtain a model of A370 
as a binary merger in Section~\ref{S:SIMULATIONS}.
Section~\ref{S:RESULTS} presents our results from hydrodynamical modeling of A370.
Section~\ref{S:Summary} contains our summary.

We adopt a spatially flat, cosmological-constant and dark-matter dominated (\LCDM) 
cosmology with $h = 0.7$, $\Omega_\mathrm{m} = 0.3$, and $\Omega_\Lambda = 0.7$.
We use the standard notation $M_\Delta$ to denote the mass enclosed within a sphere of radius $R_\Delta$,
within which the mean overdensity equals $\Delta$ times the cosmic
critical density $\rho_\mathrm{c}(z) $
at the cluster redshift $z$.
We adopt the virial overdensity $\Delta\simeq 128$
(i.e., $M_\mathrm{vir}\simeq M_{128}$) based on the spherical collapse model
(see Appendix A of \citealt{Kitayama1996}).
The quoted errors represent the 68\percent confidence level, 
unless stated otherwise.

\section{Multi-wavelength Observations of A370}
\label{S:Observations}

\subsection{X-ray observations}
\label{SS:XRData}

High angular resolution ($\sim\,$1\ASEC) X-ray observations provide
critical constraints on \NBODYHYDRO simulations because the X-ray
emission traces the intracluster gas. 
In principle, thermal SZ observations could also be used to
map the intracluster gas structure.
Although recent SZ observations are reaching the necessary angular
resolution (e.g., ALMA), spatially resolved SZ observations are limited
to a handful of clusters.
Therefore, in practice, X-ray observations are used for simulations to 
constrain the gas distribution (e.g., \citealt{Molnar2016}).

Since X-ray observations are essential to set up our simulations,
we have reanalyzed the publicly available \CHANDRA observations of A370.
The cluster was observed with 
Advanced CCD Imaging Spectrometer (ACIS)
ACIS-S (ObsID = 515) and ACIS-I (ObsID = 7715) 
with exposure times of 90 ksec and 7 ksec. 
The first observation with a long exposure time was performed in Cycle 1, 
when the focal plane temperature was warmer than the standard
$-119.7^{\circ}$\,C
The Charge Transfer Inefficiency (CTI) 
correction is valid for a focal-plane temperature of
$\sim -120^\circ$\,C.
Charge transfer inefficiency degrades the spectral resolution of the instruments.
Unfortunately no CTI correction is available for Cycle 1 observations, thus this correction
cannot be performed on these ACIS-S observations 
(e.g., \citealt{BotteonET2018MNRAS476}).
Therefore, we do not use X-ray temperature information, only morphology as a constraint
on our \FLASH simulations.

We followed standard data reduction and cleaning procedures using the updated 
versions 4.9 of Chandra Interactive Analysis of Observations 
(CIAO; \citealt{FruscioneET2006}) and version 4.7.8 of the calibration database 
(CALDB; for details of our procedure see \citealt{UedaET2019}). 
After cleaning the data, the remaining exposure time was 70\,ksec.
The background data were extracted from the region between 3.4\AMIN and
6.0\AMIN from the X-ray peak position (marking the cluster center) 
only on ACIS-S3 chip.
The Galactic absorption was fixed at $3.01 \times 10^{20}$\,cm$^{-2}$
\citep{Kalbera2005}.

\begin{figure}[t]
\includegraphics[width=0.476\textwidth]{\FIGURES/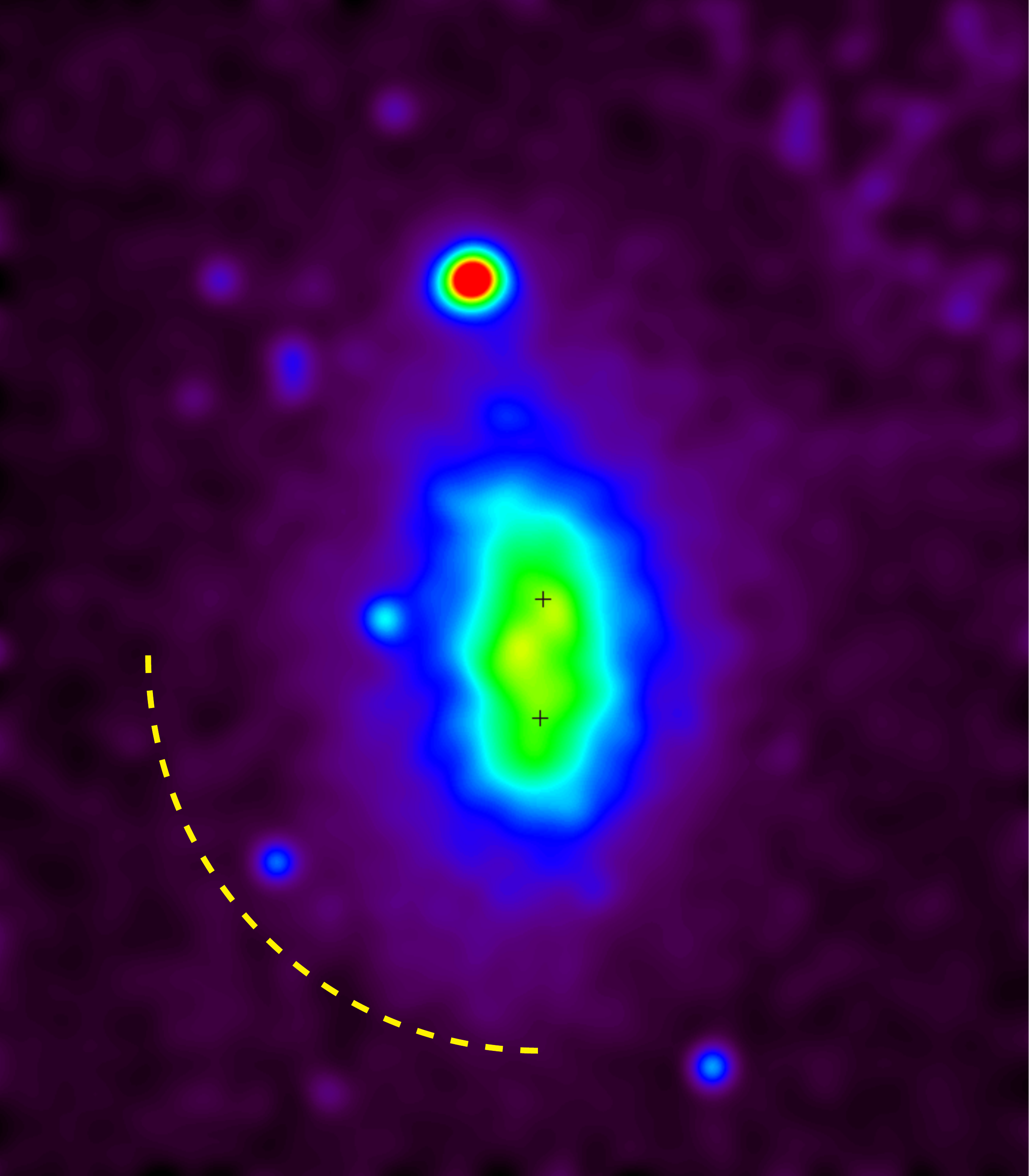}
\caption{
Smoothed X-ray surface brightness image of A370 based on \CHANDRA ACIS data.
The bright X-ray source in the north is a foreground elliptical galaxy.
The two black crosses correspond to the peak positions of dark-matter halos 
derived from strong gravitational lensing \citep{StraitET2018ApJ868}.
The image size is 1.8\,Mpc $\times$ 2.0\,Mpc. 
The yellow dashed line represents the surface brightness edge found in
 this study, in agreement with \cite{BotteonET2018MNRAS476}.
\vspace{0.05 cm}
\label{F:A37ACIS}
}
\end{figure}

In Figure~\ref{F:A37ACIS}, we show the resulting
exposure-corrected and background-subtracted X-ray surface brightness
distribution (see also the first panel in Figure~\ref{F:XRACISIMBEST}).
We smoothed the X-ray image using a Gaussian with a constant width
(5\ASEC)
 because adaptive smoothing may bring out non-existing structures in low surface 
brightness regions.
The point source in the north is a foreground X-ray bright elliptical galaxy at $z = 0.044$.
The two black crosses mark the peak positions of the two dark-matter halos 
derived from a strong-lensing analysis \citep{StraitET2018ApJ868}.

%
%

The X-ray surface brightness distribution is strongly elongated in the north-south
direction, showing a disturbed morphology with two peaks (Figure~\ref{F:A37ACIS}).
The brightest X-ray peak is located about half way between the two mass centers, 
eastward of the lines connected them at a distance of $\sim\,100$\,kpc.
The northern X-ray peak is located south-west of the northern mass peak by about 30\,kpc.
Disturbed X-ray morphology and offsets between X-ray and mass peaks are 
clear signs of dynamical activity.
As demonstrated in \cite{MolnarET2012ApJ748}, these offsets can be used to constrain 
the dynamical state of a merging cluster.

Fitting two broken power-law models to the X-ray surface brightness of A370, 
we found a drop in the X-ray emission at 690$_{-30}^{+60}$\,kpc to the east of the 
cluster center (yellow dashed line in Figure~\ref{F:A37ACIS}),
in agreement with \cite{BotteonET2018MNRAS476}, 
but found no significant edge on the west side.

\subsection{Radio Observations}
\label{S:RadioData}

We carried out a series of SZ observations of 45 galaxy clusters in the Bolocam 
X-ray SZ (BOXSZ) sample using Bolocam at 140\,GHz at the Caltech Submillimeter Observatory
\citep{CzakonET2015}. 
We found an integrated SZ amplitude within $R_{2500}$ 
of $Y_{2500} = (0.91\pm0.16) \times 10^{-10}$\,ster in A370
(see Table 3 of \citealt{CzakonET2015}).
Since the angular resolution of Bolocam at 140\,GHz is 45\ASEC, we use
only the amplitude of the SZ signal. For morphology, we use the \CHANDRA
image to constrain our \FLASH simulations.

%
%
\begin{figure*}[t]
\includegraphics[width=0.49\textwidth]{\FIGURES/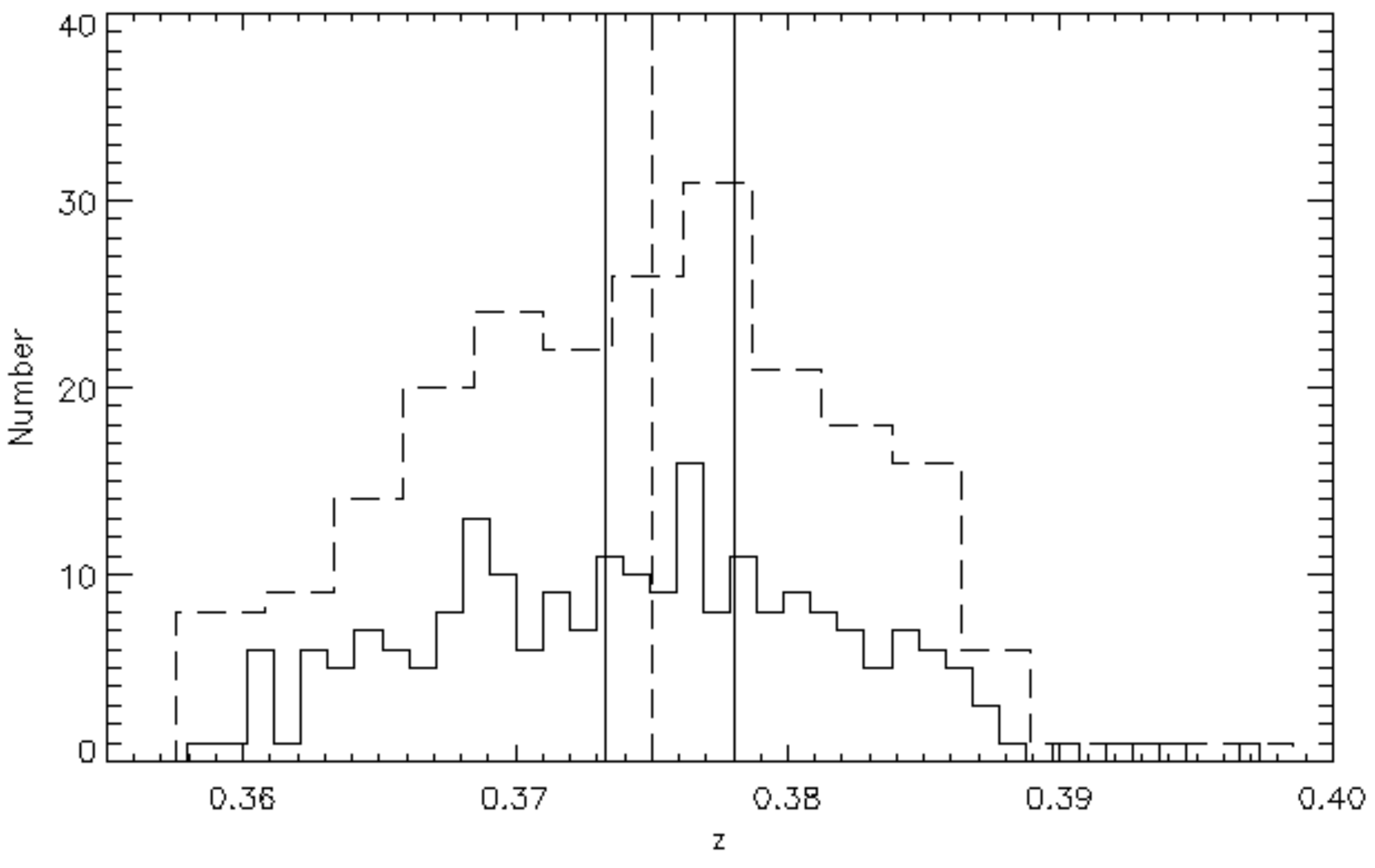}
\includegraphics[width=0.48\textwidth]{\FIGURES/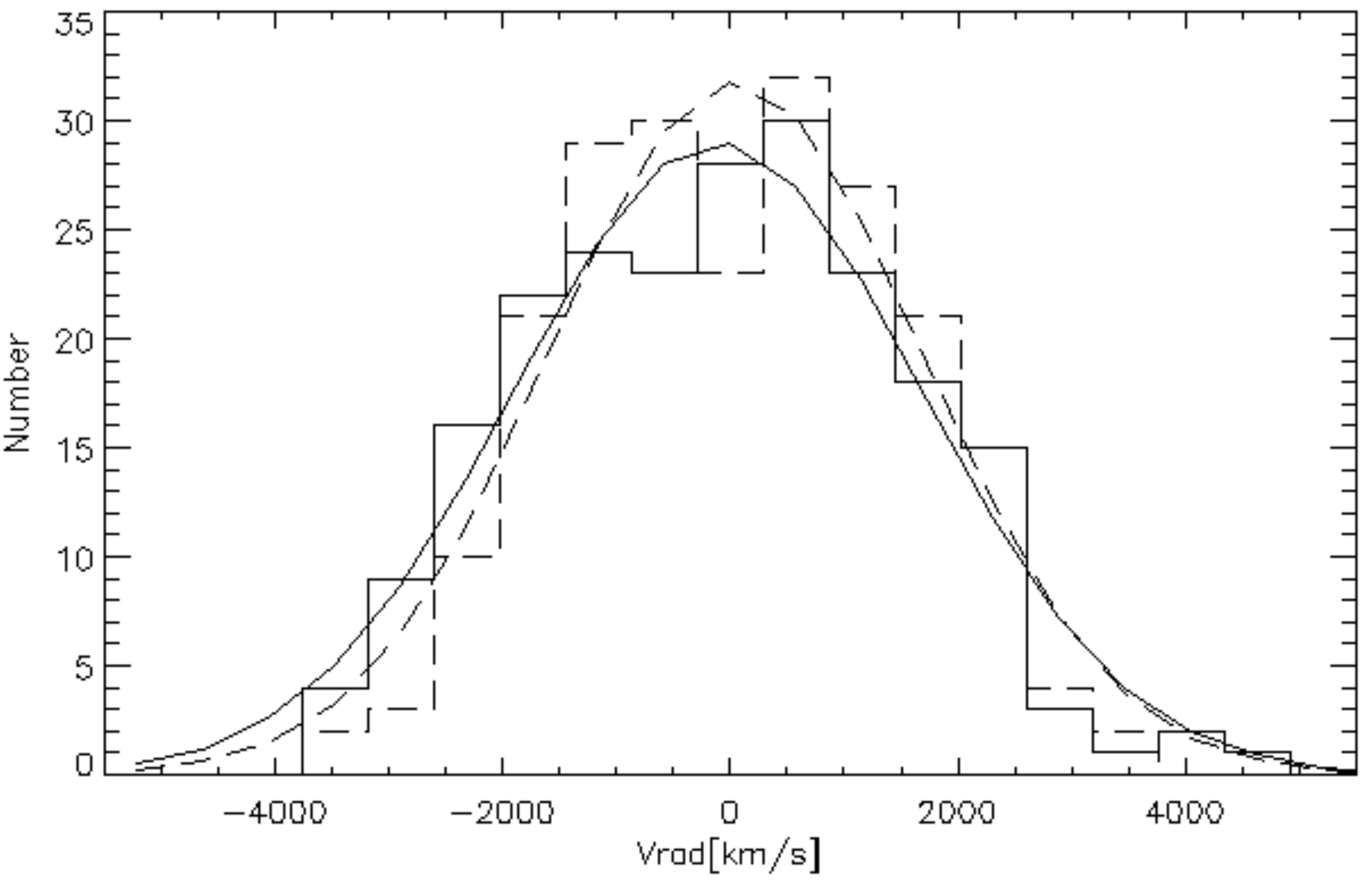}
\caption{
Left panel:
Redshift distribution of cluster member galaxies in A370 from \cite{LagattutaET2019}
using different bin sizes (solid and dashed histograms).
The cluster redshift (0.3750) and the redshifts of the two BCGs (0.3733 and 0.3780).
are marked with dashed and solid vertical lines.
Right panel: 
Relative radial velocity distribution in A370 (histograms) and Gaussian fits (curves) 
from observations and simulations (solid and dashed lines).
\vspace{0.05 cm}
\label{F:A370REDSHIFTS}
}
\end{figure*} 

\subsection{Optical Observations}
\label{S:OpticalData}

\citet{Umetsu2011a} presented a wide-field weak-lensing analysis of 5
superlens clusters (A1689, A1703, A370, Cl0024$+$17, RXJ1347$-$11) based
on Subaru telescope observations.
By combining the Subaru weak-lensing
constraints with their strong-lensing mass profile derived from
{\em HST} data \citep[see also][]{Umetsu2011b}, 
\cite{Umetsu2011a} obtained a virial mass of
$M_\mathrm{vir} = 3.25_{-0.314}^{+0.364} \times 10^{15}M_\odot$
and a concentration of $5.68\pm 0.48$
for A370 (see their Table 6),
assuming a generalized form of the Navarro--Frenk--White density profile
\citep[][hereafter NFW]{NFW1997}.

In agreement with earlier studies, recent strong-lensing reconstructions 
of the mass surface density distribution and critical curves in the cluster 
core reveal a symmetric bimodal mass distribution elongated in the
north-south direction  
\citep{StraitET2018ApJ868,DiegoET2018MNRAS473,LagattutaET2019}.
The critical curves derived recently are quite similar 
(Figure 5 of \citealt{StraitET2018ApJ868}) and compatible 
with the assumption that A370 is a merging system of two massive clusters.
However, these strong-lensing studies identified different numbers of
dark matter components in the central region,
one \citep{DiegoET2018MNRAS473}, two \citep{StraitET2018ApJ868},
or three \citep{LagattutaET2019}.
Only two BCGs were found in the center of A370, which suggests that A370
is a binary merger.  
\cite{StraitET2018ApJ868} used a method combining weak and strong lensing,
finding two main mass peaks and a few less significant peaks that are
likely associated with substructure.
Thus, we assume here that A370 is a massive binary merger and adopt a
projected distance between the two mass centers of
$D_\mathrm{proj} = 206$\,kpc \citep{StraitET2018ApJ868}.
The peak positions of two main dark matter halos (north and south)
were precisely constrained by strong-lens modeling based on a large set 
of multiple images of background galaxies,
albeit model dependent. 
We thus consider only a $\pm 30\,$kpc change in $D_\mathrm{proj}$.
The symmetry in the mass distribution implies that the ratio of the total 
masses of the two colliding clusters was close to unity, $M_1/M_2 \sim 1$.

Based on optical spectroscopy of a few cluster member galaxies, a relative 
radial velocity between the northern and southern components was 
estimated to be $V_\mathrm{rad} \sim3000$\,\KMSEC \citep{RichardET2010}. 
However, the redshifts of the northern and southern BCGs 
were found to be $z_\mathrm{N} = 0.3780$ and $z_\mathrm{S} = 0.3733$, 
respectively, which implies a relative velocity of only $V_\mathrm{rad}
\sim\,1024$\,\KMSEC at the redshift of A370 \citep{LagattutaET2019}. 
It is unlikely that the relative velocity of the BCGs to the dark-matter
center of their respective host cluster would be larger than $\sim 1000$\,\KMSEC.
In the left panel of Figure~\ref{F:A370REDSHIFTS},
we show the redshift distribution of 219 cluster member galaxies 
based on the latest survey \citep{LagattutaET2019}. 
We found no significant peaks around the redshifts of the two BCGs
($0.3733$ and $0.3780$; vertical solid lines), regardless of the chosen
bin size (see the solid and dashed histograms in
Figure~\ref{F:A370REDSHIFTS}).
The dashed vertical line marks the redshift of the cluster ($0.3750$).
In the right panel of Figure~\ref{F:A370REDSHIFTS},
we display the distribution of relative radial velocities of the cluster
members (solid histogram) and that from our best model (dashed
histogram; see Section~\ref{S:RESULTS}). 
The corresponding Gaussian fit to the observed velocity distribution
is shown with the solid curve.
The dispersion of the observed radial velocity distribution from a
Gaussian fit is $\sigma_\mathrm{obs} \approx 1795\,$\KMSEC.
Since we do not find any peaks around the redshifts of the BCGs,
we adopt $V_\mathrm{rad} = 1024\,$\KMSEC,
the relative radial velocity of the two BCGs,
and assume a large uncertainty of $\pm 30\percent$ in the relative velocity
of the two cluster components.

\section{Modeling A370 using \FLASH}
\label{S:SIMULATIONS}

\subsection{Simulation Set Up}
\label{SS:ICOND}

We model A370 in 3D using an Eulerian \NBODYHYDRO
code \FLASH developed at the Center for Astrophysical Thermonuclear Flashes
at the University of Chicago \citep{Fryxell2000ApJS131p273,Ricker2008ApJS176}.
\FLASH is a publicly available Adaptive Mesh Refinement (AMR) code, 
which can be run in parallel computer architectures based on CPUs.
We include dark matter and gas self-consistently taking the gravity of both 
components into account.

We modeled A370 as a binary merger assuming spherical symmetry for the initial 
components. 
Since the initial memory of the halo asphericity will be completely
lost after two core passages, we do not expect to be able to recover the initial 
asphericity of dark matter halos.
We adopted a box size of 19.4\,Mpc on a side. 
The highest resolution, 19\,kpc, was reached at high density regions (cluster centers), 
contact discontinuities, merger shocks, and in turbulent regions
(turbulence usually sets the refinement to its maximum in hydrodynamical simulations).
We chose a 3D Cartesian coordinate system, $x,y,z$, with the main 
plane of the collision (containing the centers of the clusters 
and the initial velocity vectors of the infalling cluster) placed in the $z = 0$ plane.
For any given relative velocity vector, we assign velocities 
of the two components to set the total momentum of the system to zero to keep 
the clusters in the simulation box. This was important because we ran our simulations 
past the third core passage, which typically takes several giga years. 
The initial velocities of the main (more massive) and the infalling (less massive) 
cluster were parallel to the $z$-axis, pointing to the $-z$ and $+z$ directions, respectively.
Our simulations were semi-adiabatic; only the most important non-adiabatic 
process in merging clusters, shock heating, was included using a shock capturing 
scheme (included in \FLASH). 
Other non-adiabatic effects, such as radiative cooling and heating 
(e.g., supernova and AGN feedback), may be safely ignored because the 
cooling time is very long in the intracluster gas in our simulated
clusters and the energy  input from heating is insignificant relative to
the energetics of the collision.

The initial models of the colliding clusters were assumed to have spherical geometry 
with cut offs of the dark-matter and gas density at the virial radius, $R_\mathrm{vir}$.
We assumed an NFW profile for the radial dark-matter distribution,
\begin{equation}
 \label{E:NFW}
 \rho_\mathrm{DM} (r) = \frac{\rho_\mathrm{s}}{(r/r_\mathrm{s})(1+r/r_\mathrm{s})^2},
\end{equation}
%
where $\rho_\mathrm{s}$ is the characteristic density parameter,
$r_\mathrm{s}=R_\mathrm{vir}/c_\mathrm{vir}$ is the characteristic
radius at which $d\ln{\rho_\mathrm{DM}}(r)/d\ln{r}=-2$,
with $\Cvir$ the concentration parameter. 
The gas density was assumed to have a $\beta$-model profile, 
\begin{equation}
\label{E:BETAMODEL}
 \rho_\mathrm{gas}(r) =
\frac{\rho_0}{[1+(r/r_\mathrm{core})^2]^{3\beta/2}}
\end{equation}
%
where $r_\mathrm{core}$ is the core radius,
$\rho_0$ is the central density parameter,
and $\beta$ determines the fall off of
the gas density at large radii. We assumed $\beta=1$, as suggested by 
cosmological simulations for the large scale distribution 
of the intracluster gas in relaxed clusters (derived excluding filaments; 
for details see \citealt{Molnet10ApJ723p1272}). 
We used a gas mass fraction of  $f_\mathrm{gas} = 0.12$
\citep[e.g.,][]{Vikhlinin2006,Umetsu2009,Tian2020}, 
and assumed, as an approximation, that the galaxies are collisionless
and follow the dark-matter distribution.

We determined the initial gas temperature distribution as a function of
the cluster-centric radius, $T(r)$, assuming \HE\
and $\gamma = 5/3$ for the ideal gas equation of state.
It is more difficult to model a stable dark-matter density distribution.
The velocity field has to be set up initially to provide the required stable 
density distribution dynamically as the dark-matter particles move around
on their orbits.
With the assumptions of spherical density distribution and isotropic velocity dispersion, 
the Jeans equation can be solved for the amplitude of the velocity dispersion,
$\sigma_v (r)$, as a function of $r$ 
\citep{LokasMamon2001}.
We obtain the amplitudes of the velocities for dark-matter points at radius $r$ 
by sampling the $\sigma_v (r)$ distribution derived from the Jeans equation 
and choose an angle from an isotropic distribution,
which is referred to as the local Maxwellian approximation.
For further details of the set up of our simulations, see \cite{MolnarET2012ApJ748}.

%
%
\begin{deluxetable}{lcccccc}[t]
\tablecolumns{11}
\tablecaption{                       \label{T:TABLE1} 
 Input parameters used in our \FLASH simulations.\\
} 
\tablewidth{0pt} 
\tablehead{ 
 \multicolumn{1}{l}		{ID\,\tablenotemark{a}}                  &
 \multicolumn{1}{c}   		{$M_{1}$\,\tablenotemark{b}}     &
 \multicolumn{1}{c} 		{$c_1$\,\tablenotemark{c}}             &
 \multicolumn{1}{c}  		{$M_{2}$\,\tablenotemark{b}}     &
 \multicolumn{1}{c} 		{$c_2$\,\tablenotemark{c}}             &
 \multicolumn{1}{c} 		{$P$\,\tablenotemark{d}}                   &
 \multicolumn{1}{c}  	  	{$V_\mathrm{in}$\,\tablenotemark{e}}        \\
         & $10^{15}M_\odot$ &  $10^{15}M_\odot$  &  &    & kpc & km\,s$^{-1}$
 }
 \startdata  
   RP100V3.0      &   1.7    &   7.0  &   1.6   &   7.0  &   100   &   3000   \\ \hline  
   RP100V3.5      &   1.7    &   7.0  &   1.6   &   7.0  &   100   &   3500   \\ \hline  
   RP100V3.7      &   1.7    &   7.0  &   1.6   &   7.0  &   100   &   3700   \\ \hline  
   RP100V4.0      &   1.7    &   7.0  &   1.6   &   7.0  &   100   &   4000   \\ \hline  
   RP200V3.5      &   1.7    &   7.0  &   1.6   &   7.0  &   200   &   3500   \\ \hline  
   RP200V3.0      &   1.7    &   7.0  &   1.6   &   7.0  &   200   &   3000   \\ \hline   
   RP300V3.0      &   1.7    &   5.0  &   1.6   &   5.0  &   300   &   3000   \\ \hline   
   RP400V2.0      &   1.7    &   5.0  &   1.6   &   5.0  &   400   &   2000   \\ \hline   
   RP400V3.0      &   1.7    &   6.0  &   1.6   &   7.0  &   400   &   3000   \\ \hline   
   RP100V3.5a    &   1.7    &   5.0  &   1.6   &   5.0  &   100   &   3500   \\ \hline   
   RP100V3.5b    &   1.7    &   6.0  &   1.6   &   6.0  &   100   &   3500   \\ \hline  
   RP100V3.5c    &   1.7    &   8.0  &   1.6   &   8.0  &   100   &   3500                   
\enddata
\tablenotetext{a}{IDs of the runs}
\tablenotetext{b}{Virial masses $\Mvir$ of the main and the infalling cluster ($M_1$ and $M_2$)}
\tablenotetext{c}{Concentration parameter $\Cvir$ for each component ($c_1$ and $c_2$).}
\tablenotetext{d}{Impact parameter.}
\tablenotetext{e}{Infall (relative 3D) velocity.
\vspace{.4 cm}}
\end{deluxetable}  

\subsection{\FLASH Simulations}
\label{SS:RUNS}

The large masses of the two cluster components and the long time
to reach the third core passage make a systematic search in the full
parameter space  
(i.e., the initial masses, concentration parameters,  infall velocity,
and impact parameter)
unfeasible using currently
available conventional high-speed computing nodes based on CPUs.
Therefore, to reduce the computer demand, 
we inspected simulations from our extensive runs of 
existing simulations and performed new simulations
by reducing the parameter space to find a reasonable model for A370.

Taking advantage of the north-south symmetry of the 
mass distribution based on gravitational lensing,
which suggests that the mass ratio $M_1/M_2$ is close unity
(Section~\ref{S:OpticalData}), we fixed the virial masses at
$M_1 = 1.7 \times 10^{15}M_\odot$ and $M_2 = 1.6 \times 10^{15}M_\odot$.
Here we break the unphysical perfect symmetry by choosing $M_1/M_2 \neq 1$.
A slight change of the total mass, or mass ratio, would not cause 
significant change in the projected X-ray morphology and other
parameters, while the best epochs and view angles would be somewhat
different. 
By inspecting the merging cluster simulations in our extensive 
data base and performing additional simulations, 
we find that significantly different initial masses do not provide 
a good match with the observations.
Fixing the initial masses allows us to reduce the parameter space
to a manageable level. 
We carried out a series of \FLASH simulations varying the impact parameter, 
the infall velocity, and the concentration parameters of our model. 
Our aim was to find a physical model for A370 with 
a reasonable agreement with the multi-wavelength observations.

We used the following constraints to find the best model for A370:
(i) Peak offset in the projected dark matter distribution
$D_\mathrm{proj} = 206$\,kpc \citep{StraitET2018ApJ868}:
since the uncertainty in $D_\mathrm{proj}$, $\sim 30$\,kpc,
results only in a small change in the rotation angles
and the noise fluctuations in the mock \CHANDRA images are larger than 
the change in the X-ray surface brightness,
we kept  $D_\mathrm{proj}$ fixed at this value;
(ii) Relative radial velocity $V_\mathrm{rad} = 1024\,$\KMSEC:
we allowed a $\pm 30\percent$ variation in $V_\mathrm{rad}$
(Section~\ref{S:OpticalData});
(iii) SZ amplitude $Y_{2500} = 0.91 \times 10^{-10}$\,ster 
based on our Bolocam measurements \citep{CzakonET2015}, allowing a
$2\sigma$ offset (Section~\ref{S:RadioData});
(iv) X-ray morphology: 
we focused on the positions of the X-ray peaks and their offsets from the 
mass peaks because of the low counts in the outer regions. 
The location of the main X-ray peak was required to lie between the two 
mass centers, about half way with a small offset to the west, with an extension 
toward the southern mass peak, and the secondary peak to lie close to the
northern mass center with a small offset toward south-west. 
Our model constraints were based on quantitative criteria, except for
those based on the X-ray morphology.

From our simulations we choose those epochs for which a viewing  
angle that satisfies all the three quantitative criteria ((i), (ii), and
(iii)) can be found. Then, the best match with observations 
was found by inspecting visually the simulated \CHANDRA images 
based on our criteria for the X-ray morphology (see the next section for a 
description of our method to generate mock \CHANDRA observations).
Since our \FLASH simulations have no feedback, heating, and cooling processes,
we use the integrated SZ amplitude, $Y_{2500}$, to constrain the physical
state of the gas, which is proportional to the dynamically important quantity, 
the thermal pressure. The pressure distribution in our simulations is more realistic 
than other gas quantities because the history of merging clusters is determined 
mostly by hydrodynamics and the balance between pressure and gravitational forces.

Our procedure of using simulations was as follows:
We used a trial-and-error approach to run simulations and find a set 
of initial conditions that satisfy our four criteria. A systematic parameter 
search was not possible because of large parameter space demanding 
too much computer time. After finding the best match, we ran additional 
simulations with a range of different initial conditions to constrain the 
parameter space.

In Table~\ref{T:TABLE1}, we summarize the initial parameters 
of our \FLASH simulations that are relevant to this paper.
This represents a small subset of parameter space that contains our best
solution and a few simulations to illustrate the effects of varying the
parameters around the values of our best model.
Here we did not include the parameters of all merging cluster
simulations because it is not informative.
The first column contains the IDs of our runs indicated as 
RP$ijk$V$nm$, with P$ijk$ the impact parameter in units of kpc
and V$nm$ the infall velocity in units of 1000\,\KMSEC.
Table~\ref{T:TABLE1} also lists the impact parameter $P$ in 
kpc and the 3D infall velocity $V_\mathrm{in}$ in \KMSEC.
The infall velocity we adopt is the relative velocity of the two clusters 
at the time when the two intracluster gas touch as they collide, 
i.e., when the distance between them is the sum of their virial radii.

%
%
\begin{figure*}[t]
\includegraphics[width=0.2376\textwidth]{\FIGURES/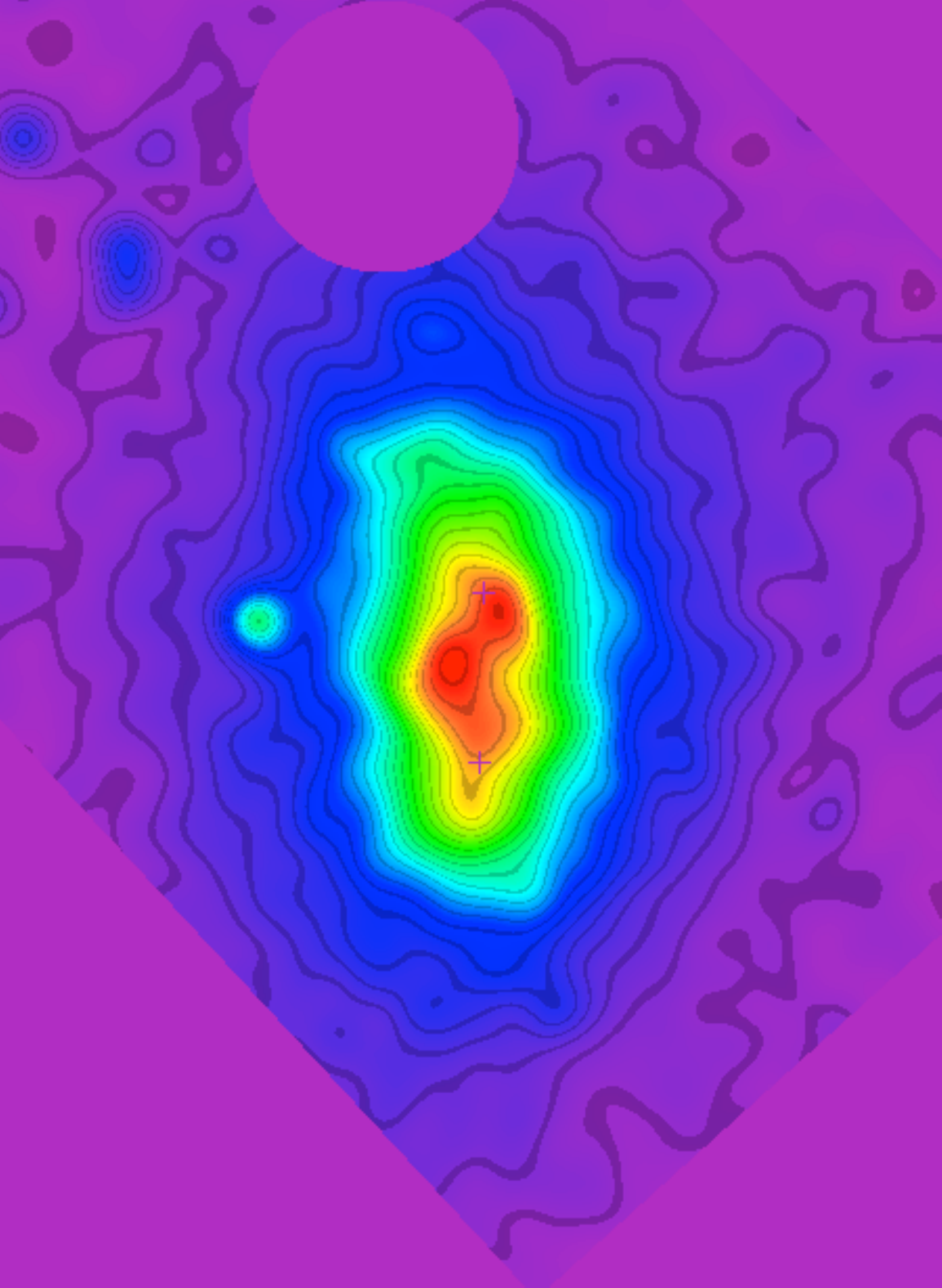}
\hspace{-0.75mm} 
\includegraphics[width=0.2376\textwidth]{\FIGURES/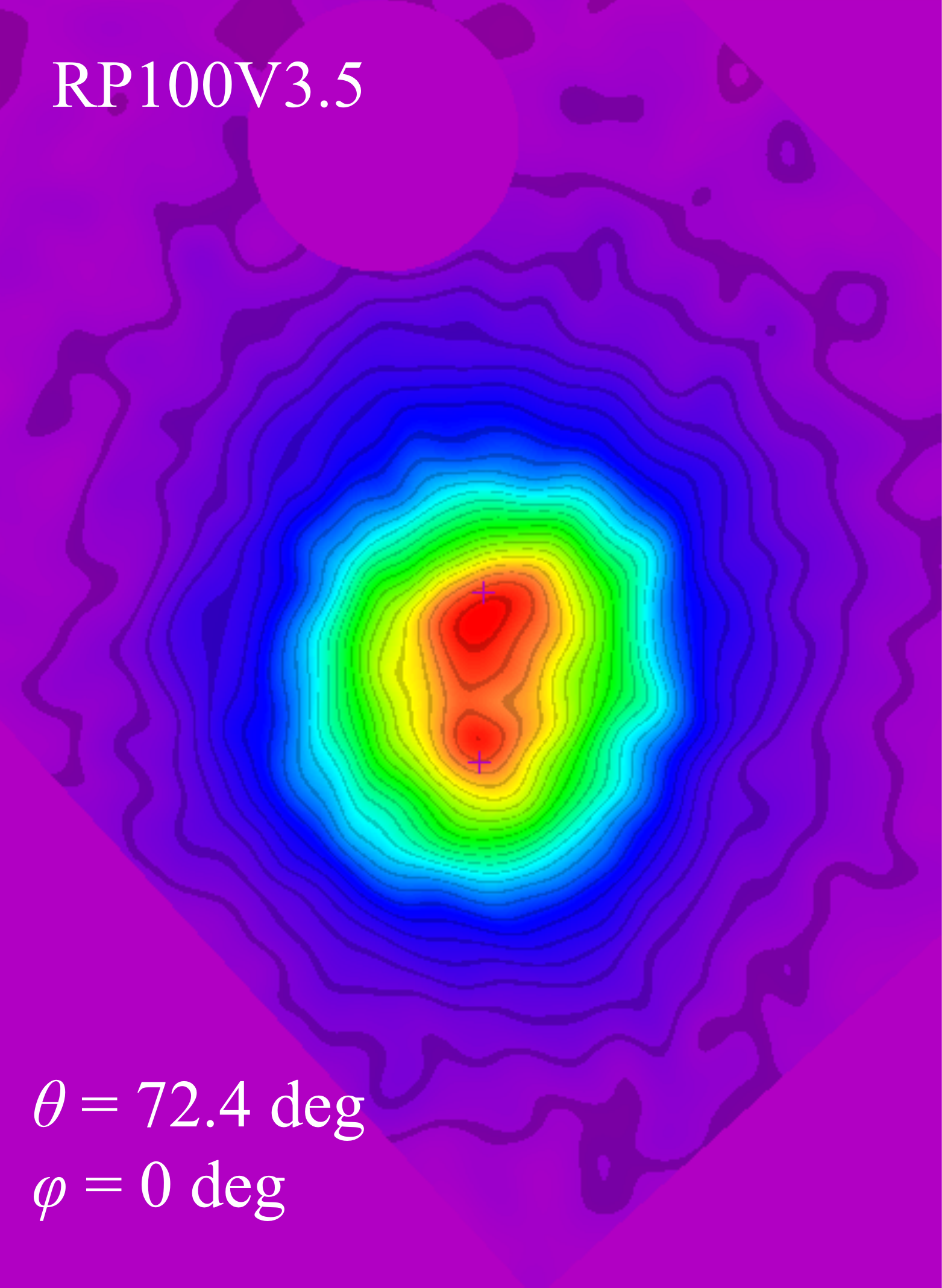}
\hspace{-0.75mm} 
\includegraphics[width=0.2376\textwidth]{\FIGURES/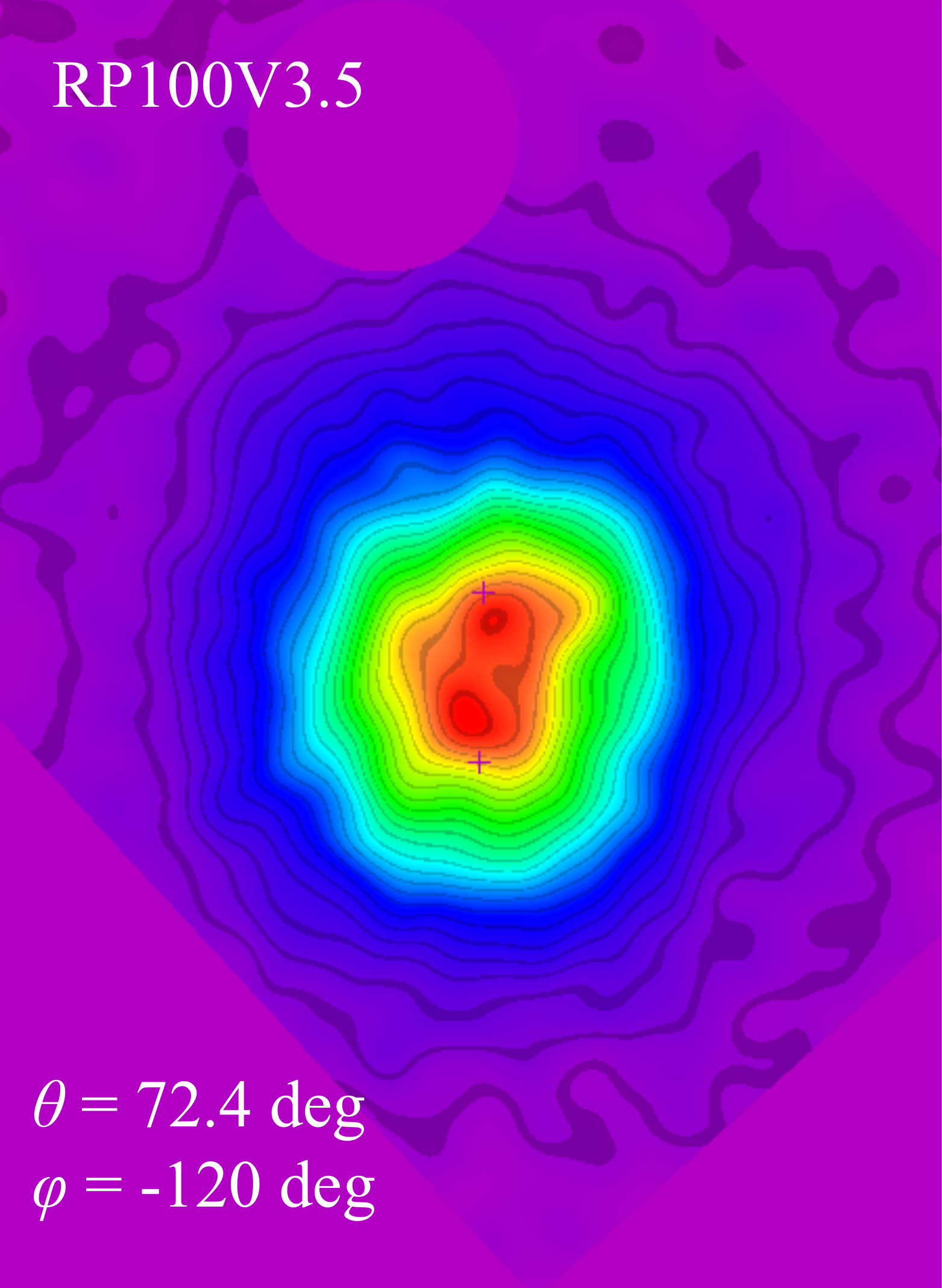}       
\hspace{-0.75mm} 
\includegraphics[width=0.2376\textwidth]{\FIGURES/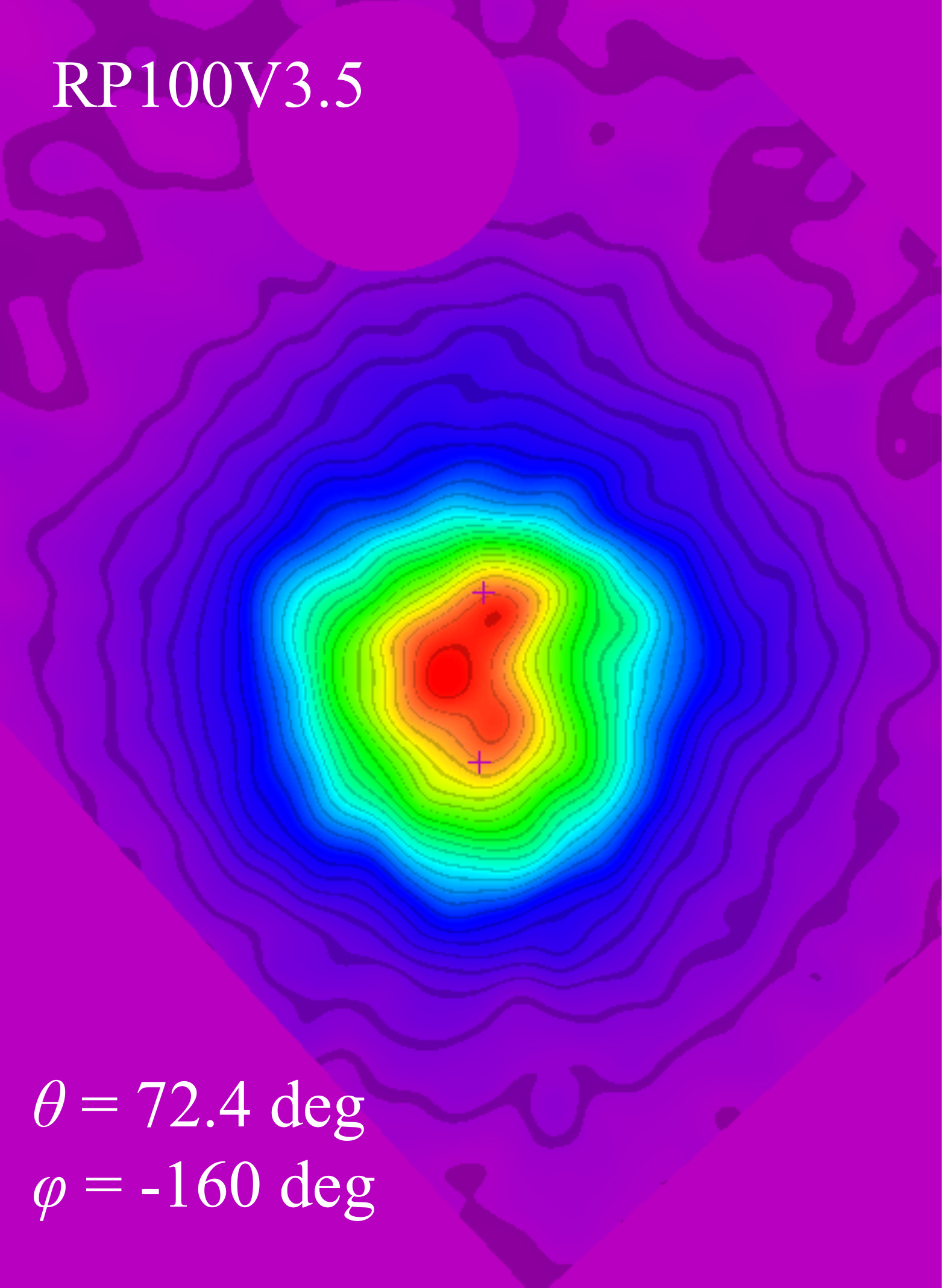}
\hspace{-1.4mm} 
\includegraphics[width=0.0359\textwidth]{\FIGURES/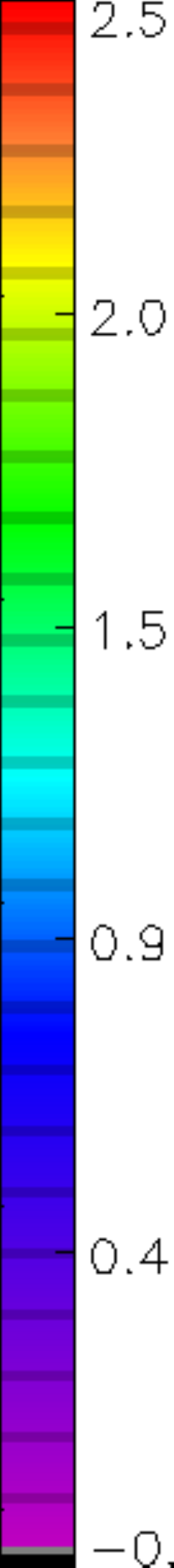}
\caption{
X-ray surface brightness images of A370 based on \CHANDRA observations
and our simulations at the best epoch and with different viewing angles.
The images are 1.14\,Mpc $\times$ 1.56\,Mpc.
The two black crosses in each panel mark the peaks of the mass surface
 density distribution.
Left to right: 
first panel: \CHANDRA X-ray image.
The area around the bright X-ray source in the north, a foreground 
elliptical galaxy, was masked in the observed image and also in the 
simulated images for easier comparison. 
Second, third, and fourth panels: X-ray images based on our best
 simulation run (RP100V3.5) at the best epoch after the second core passage,
with different roll angles ($\varphi =0^\circ, -120^\circ, -158^\circ$).
The epochs and the viewing angles were chosen 
to match the positions of the dark-matter centers, the relative radial velocities 
 of the two merging clusters with the observations,
 and, if possible, to resemble the observed X-ray morphology. 
\vspace{0.1 cm}
\label{F:XRACISIMBEST}
}
\end{figure*} 

\subsection{Monte Carlo Simulations of X-ray Images}
\label{SS:MonteCarloSIM}

We have used Monte Carlo simulations to generate mock \CHANDRA X-ray
surface brightness images. 
We defined our Cartesian coordinate system as the $z-x$ plane being
in the plane of the sky and the $z$ axis connecting the 
two dark-matter mass centers pointing to the direction of motion of the infalling cluster.
We chose the $y$ axis so that $x$, $y$, and $z$ form a right handed 
coordinate system. 
At each epoch we considered for output, we rotated the simulated cluster  
around the $z$ axis by different roll angles $\varphi$.
We then rotated the cluster out of the plane of the sky around the $y$ axis
by a polar angle $\theta$ to obtain $D_\mathrm{proj} = 206$\,kpc 
between the two dark-matter centers, matching the projected distance derived from   
gravitational lensing \citep{StraitET2018ApJ868}.
All 3D rotations were carried out using the IDL function ROT, 
which relies on an interpolation scheme to find values at the rotated pixel centers.
We derived the 3D distances between the mass peaks based on the particle 
output from \FLASH and calculated the the polar angles $\theta$
by requiring the projected distance to be $D_\mathrm{proj} = 206$\,kpc.

We generated mock images of the cluster using the same pixel size, 0.492\ASEC, 
as the ACIS detectors of \CHANDRA \citep{Gamire2003}
by integrating the X-ray emissivity along the line-of-sight (LOS) 
projected to the same sky coordinates as the observations
(the sky coordinates of the observations can be obtained from the corresponding FITS files).
We used the X-ray surface brightness image based on our analysis of \CHANDRA
data to read off the number of background photons per pixel.
We generated mock \CHANDRA images using Monte Carlo simulations based 
on adding the number of photons expected from the cluster and the background. 
A more detailed description of our methods to generate simulated 
X-ray surface brightness and mass surface density images 
can be found in \cite{MolnarBroadhurst2015}.
When comparing our mock X-ray surface brightness images to the observed ones,
we exclude ACIS chip gaps and keep only the image of the chip 
that contains the center of A370. We also excise an area around the 
foreground bright elliptical galaxy for clarity.

%
%
\begin{figure*}[t]
\includegraphics[width=0.2376\textwidth]{\FIGURES/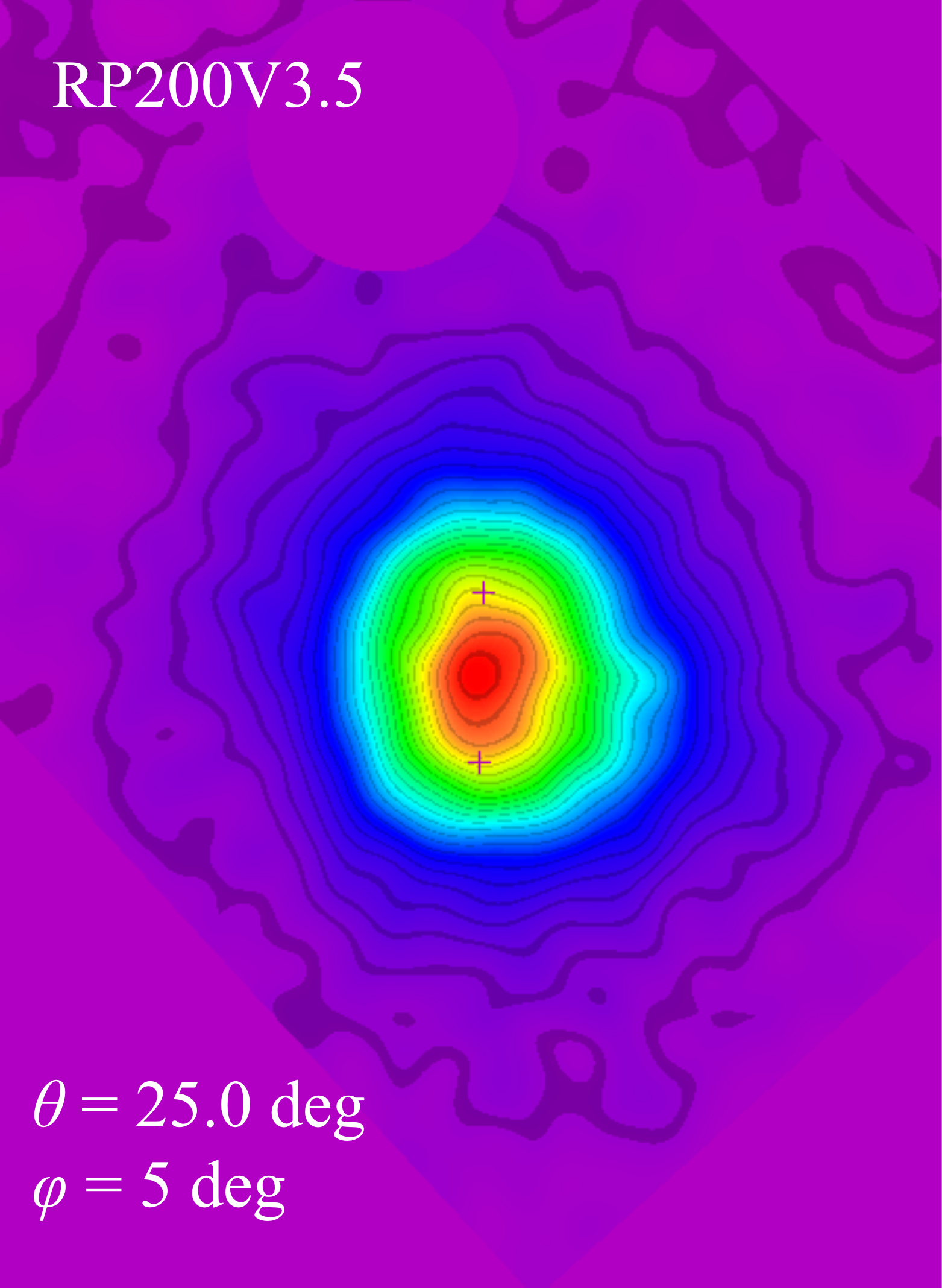} 
\hspace{-0.7mm} 
\includegraphics[width=0.2376\textwidth]{\FIGURES/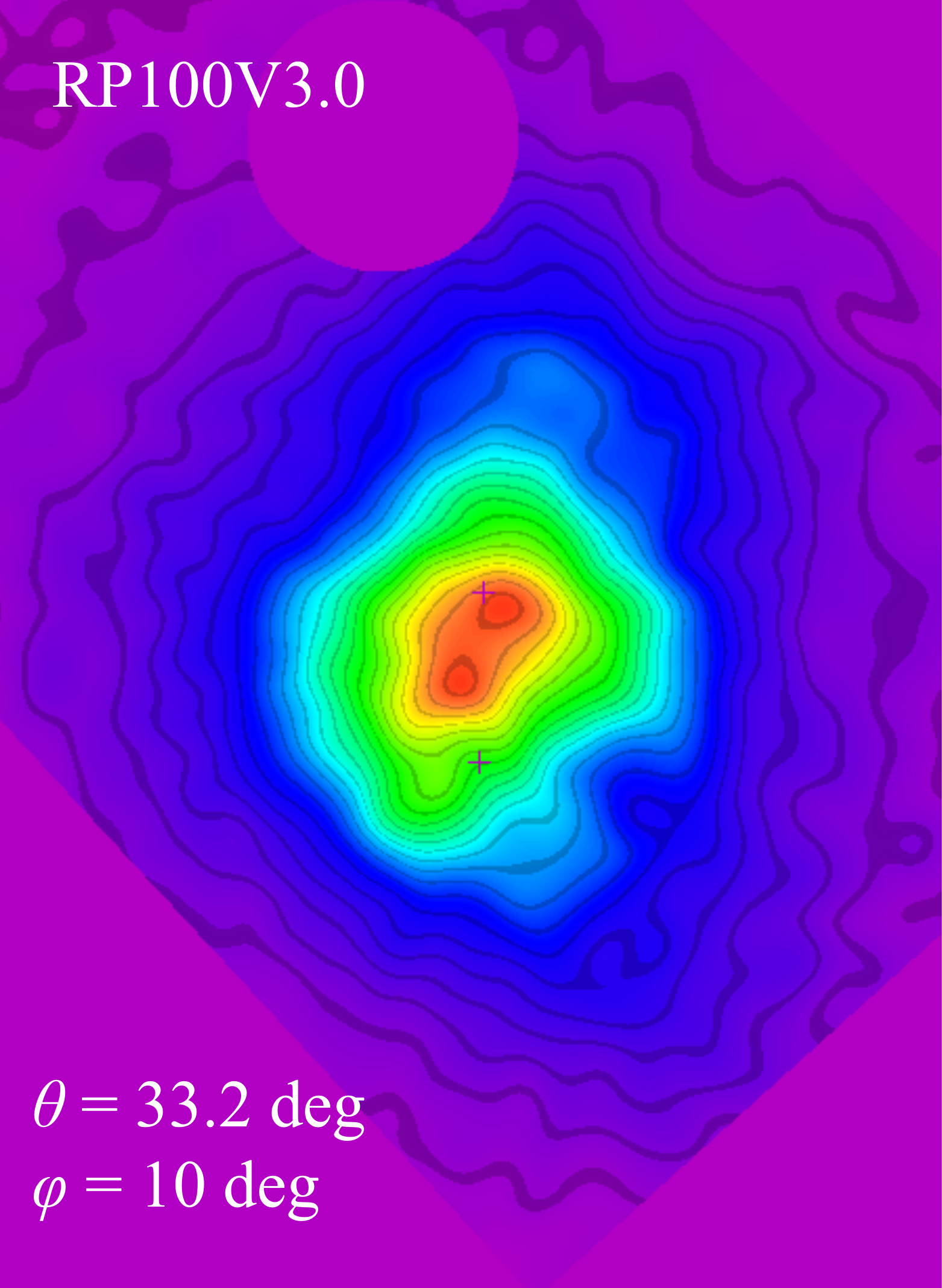}
\hspace{-0.8mm} 
\includegraphics[width=0.2376\textwidth]{\FIGURES/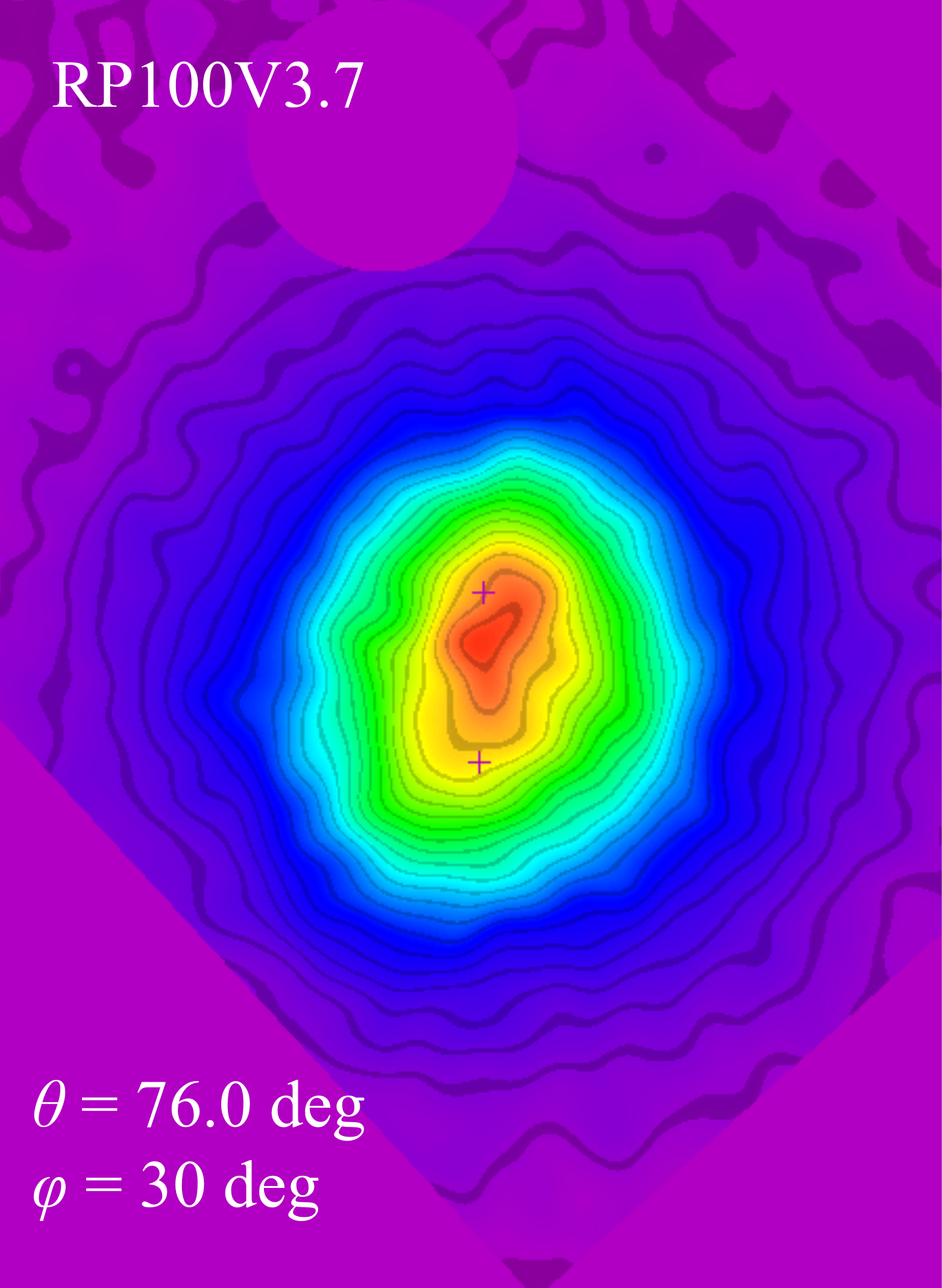} 
\hspace{-0.6mm} 
\includegraphics[width=0.2376\textwidth]{\FIGURES/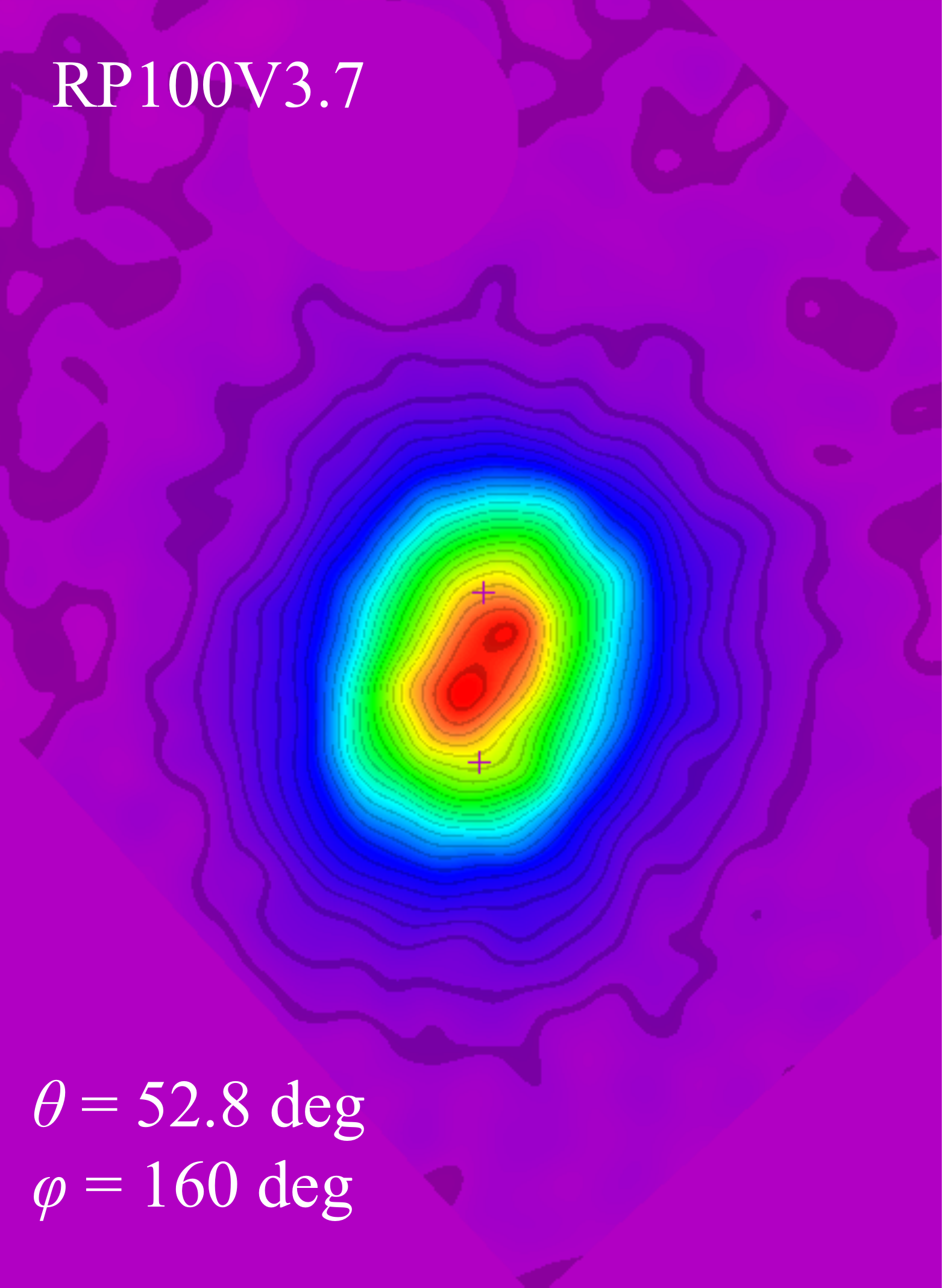}
\hspace{-1.45mm} 
\includegraphics[width=0.0359\textwidth]{\FIGURES/fig3e.pdf}	\\	
\includegraphics[width=0.2376\textwidth]{\FIGURES/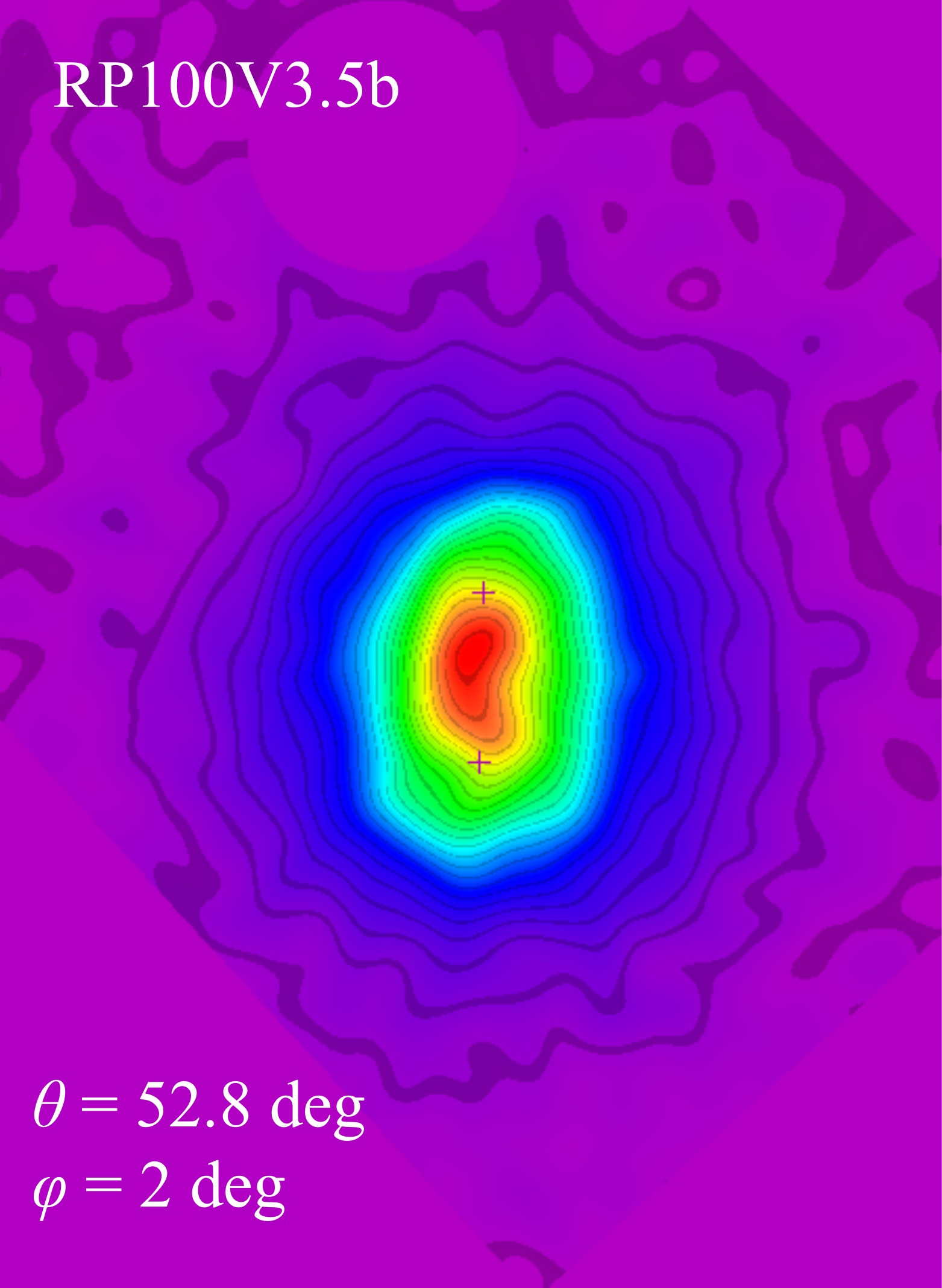} 
\includegraphics[width=0.2376\textwidth]{\FIGURES/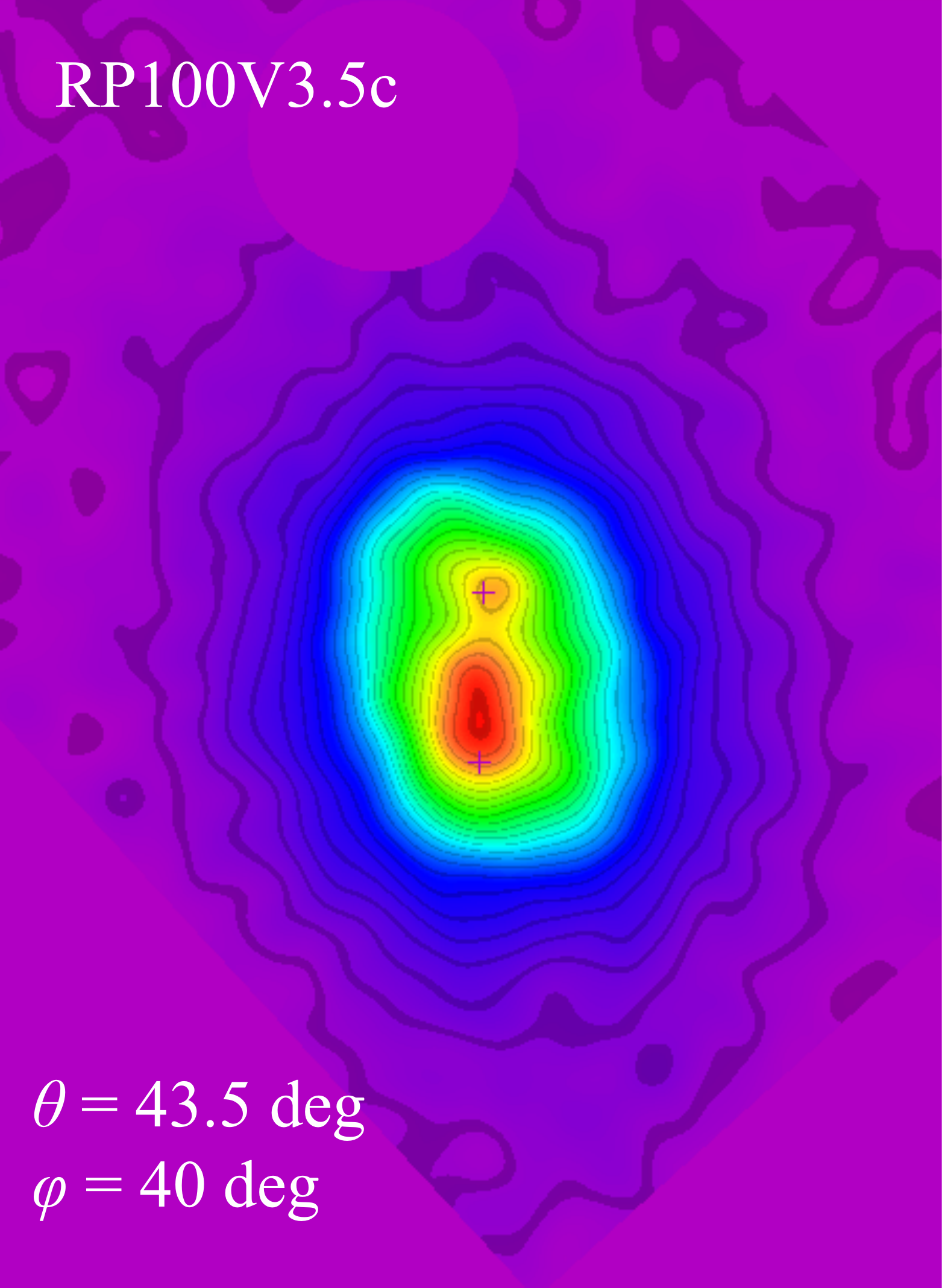}
\includegraphics[width=0.2376\textwidth]{\FIGURES/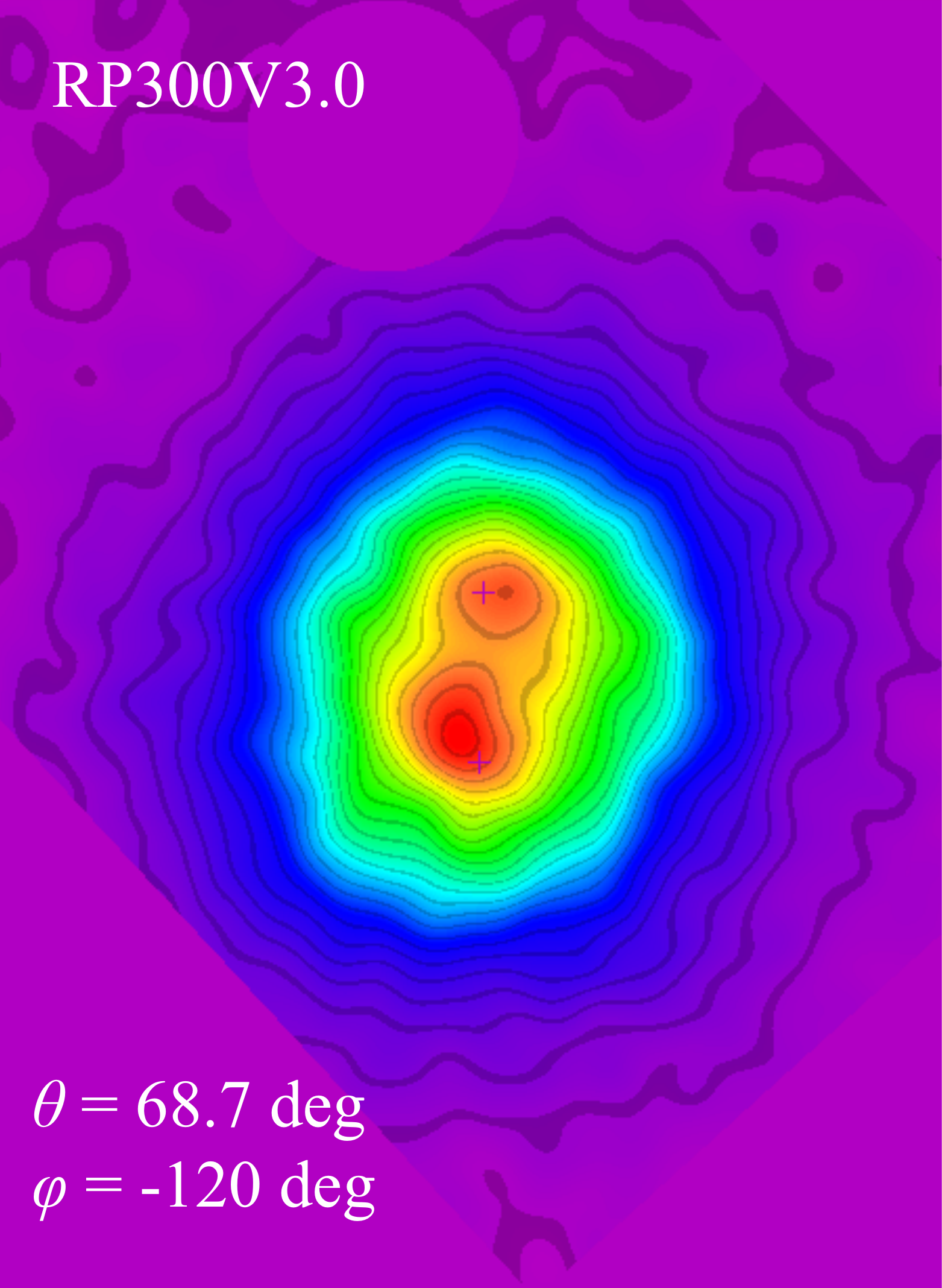}  
\includegraphics[width=0.2376\textwidth]{\FIGURES/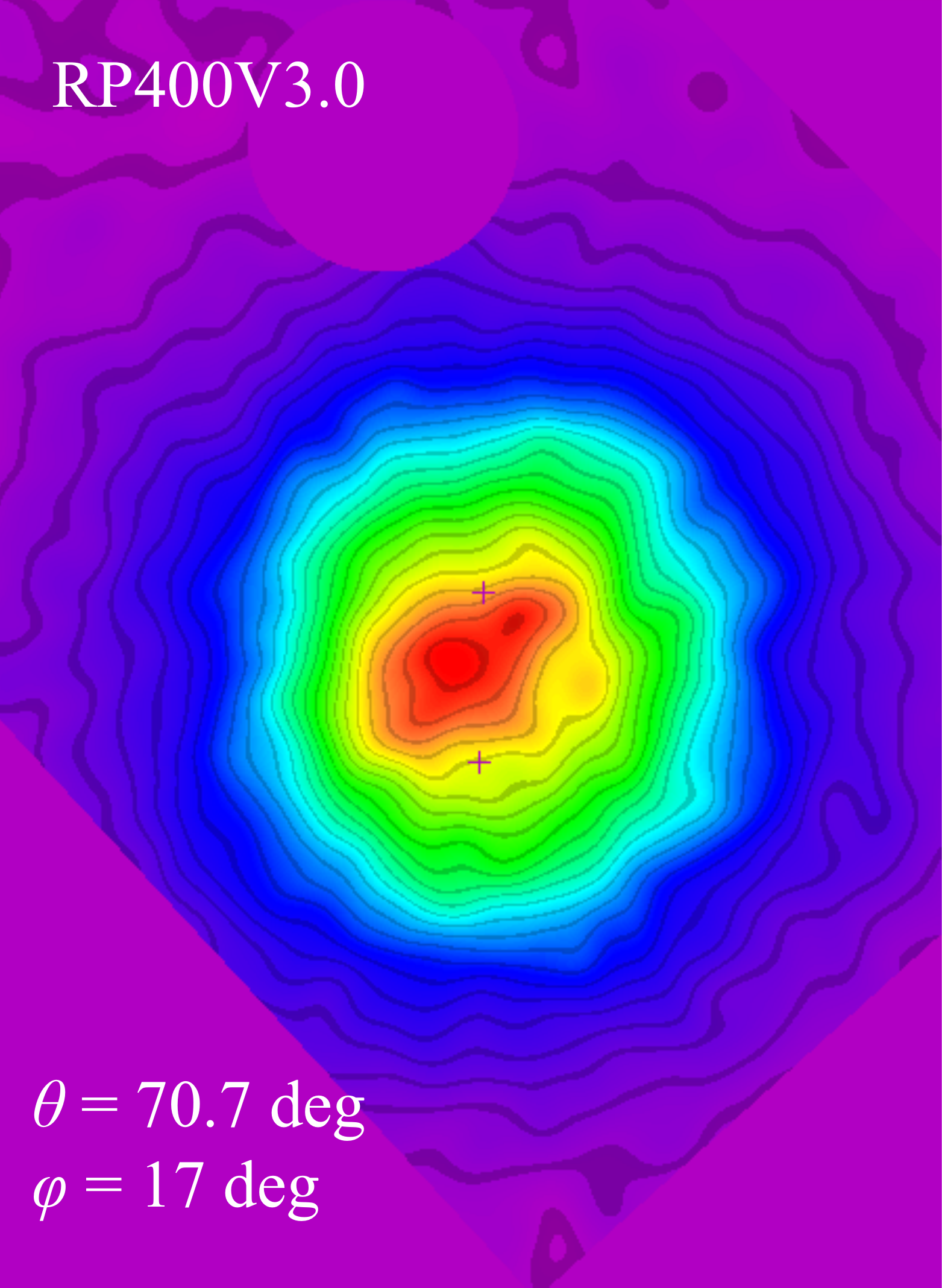}
\hspace{-1.4mm} 
\includegraphics[width=0.0359\textwidth]{\FIGURES/fig3e.pdf}		
\caption{
X-ray surface brightness images of A370 based on our simulations.
Most of the initial parameters were the same as in our best model, but we changed 
one or two parameters for each run (see Table~\ref{T:TABLE1}) to show how the 
changes impact the X-ray morphology.
The epochs and the viewing angles were chosen to match the observations
and, if possible, to resemble the observed X-ray morphology. 
From left to right:
First row:
first panel: our simulation with an impact parameter of $P=200$\,kpc (RP200V3.5);
second panel: our simulation with an infall velocity of $V_\mathrm{in}=3000$\,\KMSEC (RP100V3.0);
third and fourth panels: our simulation with an infall velocity of
 $V_\mathrm{in}=3700$\,\KMSEC (RP100V3.7) at two different epochs after
 the second core passage. 
Second row:
first panel: the concentration parameters are changed to $c_1 = c_2 = 6$
                (RP100V3.5b);
second panel: the concentration parameters are changed to $c_1 = c_2 = 8$
                (RP100V3.5c);
third panel: the impact parameter is changed to $P=300$\,kpc and the
 infall velocity to $V_\mathrm{in}=3000$\,\KMSEC 
                (RP300V3.0);
fourth panel: the impact parameter is changed to $P=400$\,kpc, the infall velocity to $V_\mathrm{in}=3000$\,\KMSEC, 
                 and the concentration parameters to $c_1 = 6$ and $c_2 = 7$
                 (RP400V3.0).
\vspace{0.1 cm}
\label{F:SIMUBAD}
}
\end{figure*} 

\section{Results and discussion}
\label{S:RESULTS}

We show in Figures~\ref{F:XRACISIMBEST} and \ref{F:SIMUBAD}
the X-ray surface brightness image of A370 derived 
from \CHANDRA data and the images of mock \CHANDRA 
observations based on our \FLASH simulations. 
Here we applied the same smoothing (5\ASEC) on the observed and simulated images
for comparison.
The images are 1.14\,Mpc $\times$ 1.56\,Mpc. 
We used the same color code for all panels and excluded the area
containing the foreground elliptical galaxy in the north for clarity.
For each mock observation, the epoch and the viewing angle were chosen
such that
the relative LOS velocity matches $V_\mathrm{rad} \sim 750$\,\KMSEC
and
the locations of the two peaks of the mass surface density are aligned
with those observed.\footnote{This can be achieved with a proper choice of the
polar angle, $\theta$, if the 3D separation is $\ge 206$\,kpc at a given epoch.}

In general, a viewing angle that produces a projected mass peak
separation of $D_\mathrm{proj} \sim 206$\,kpc
and a relative LOS velocity of $V_\mathrm{rad} \sim 750$\,\KMSEC
may be chosen at a few epochs before and after the core passages, 
as long as the 3D velocity  due to the accelerating or decelerating
infalling cluster is greater than $\sim 750$\,\KMSEC.
We found that projections with a LOS relative velocity of
$V_\mathrm{rad} \sim 750$\,\KMSEC
before/after the first core passage and before the second core passage
appear to result in a single bright peak in the X-ray surface brightness, independently 
of the infall velocity and impact parameter of the collision and the
viewing angle.
This is illustrated in Figure~\ref{F:SIMUBAD} (see the first panels in
the first and second rows). 
The relative velocities in the simulations at epochs after the third core 
passage are usually too low to produce the required relative radial velocity.
Therefore, we shall focus on simulations at epochs between the second
and third core passages.

The first panel of Figure~\ref{F:XRACISIMBEST}
shows the smoothed \CHANDRA X-ray image based on our analysis (see Section~\ref{SS:XRData}). 
Images shown in the other panels are based on our best simulation run 
(RP300V3.5) at the best epoch, 0.57\,Gyr (6.8\,Gyr) after the second (first) core passage
with a polar angle of $\theta = 72.4^\circ$
and three different roll angles, namely
$\varphi = 0^\circ, -120^\circ$, and $-158^\circ$.
The X-ray morphology with a roll angle of $\varphi =-158^\circ$ provides
the best match with the observations.
Contrary to the observations,
views with roll angles of $\varphi =0^\circ$ and $-120^\circ$ exhibit
two X-ray peaks near their associated mass peaks, but no X-ray peak 
around half way between them.

The best model is provided by our RP100V3.5 run at an epoch of 0.57\,Gyr after 
the second core passage and just before the third core passage,
with a polar angle of $\theta = 72.4^\circ$ and a roll angle of $\varphi = -158^\circ$
(see the third panel in Figure~\ref{F:XRACISIMBEST}).
The main X-ray peak is located between the two mass centers, about half way with a 
small offset to the west. It shows an extension toward the southern mass
peak and the secondary peak is close to the northern mass center, with a
small offset toward south-west. 
With this viewing angle, the two cluster components of RP100V3.5 
has a relative radial velocity of $V_\mathrm{rad} = 960$\,\KMSEC
and an integrated SZ amplitude of $Y_{2500} = 1.2 \times 10^{-10}$\,ster.

We show in the right panel of Figure~\ref{F:A370REDSHIFTS}
(dashed histogram)
the distribution of relative radial velocities constructed from 
our best model.
The corresponding Gaussian fit to our simulated histogram is
shown with the dashed curve. 
We obtain a velocity dispersion of
$\sigma_\mathrm{sim} \approx 1654$\,\KMSEC
based on our best model, which agrees well within $10\percent$ with the  
observed value 
($\sigma_\mathrm{obs} \approx 1795$\,\KMSEC; see
Section~\ref{S:OpticalData}).

We thus conclude that, our simulation (RP100V3.5) with total virial masses of  
$M_{1} = 1.7 \times 10^{15} \,M_\odot$ and
$M_{2} = 1.6 \times 10^{15} \,M_\odot$,
an initial impact parameter of $P \sim 100$\,kpc,
a 3D infall velocity of $V_\mathrm{in} \sim 3500$\,\KMSEC,
and concentration parameters of $c_1 = 6$ and $c_2 = 6$
is the best dynamical model for A370.
Our best model is in agreement with an earlier interpretation of
multiwavelength observations of A370, that is, it is a massive post
major merger with $M_1/M_2 \sim 1$. 
Moreover, the distribution of radial velocities derived from our best
model is very similar to that observed in A370
(compare the dashed and solid histograms in the left panel of
Figure~\ref{F:A370REDSHIFTS}).  
This agreement provides an independent confirmation of the large mass 
derived from weak lensing by \cite{Umetsu2011a,Umetsu2011b}.

We note that the shape of the outer X-ray brightness distribution from
our best model is rounder than the observed one, which is a consequence
of a shallower fall off of the gas density distribution than that of A370.
Here we did not aim to improve the fit of the outer regions, because it
would require an even more extensive search in an extended parameter 
space including the outer density slope of the initial clusters, which would 
increase the computer time significantly. However, on the basis of our 
extensive simulations, we do not expect a significant difference in the 
cluster core regions from changing the slope in the outer regions with 
much lower gas densities.

We have also performed a few simulations using different values of
 $P$, $V_\mathrm{in}$, and $c_\mathrm{vir}$ to derive crude
constraints on our best initial parameters (see Table~\ref{T:TABLE1} for
the list of parameters of simulations discussed in our paper). 
Specifically,
we ran simulations with the infall velocity $V_\mathrm{in}$ in the range of 
$2000$ to $4000$\,\KMSEC, the impact parameter $P$ from 100 to 400\,kpc,
and the concentration parameter $c_\mathrm{vir}$ from 5 to 8 
to cover a representative range of parameter space. 
We did not perform simulations with zero impact parameter because
that would result in an X-ray morphology with axial symmetry contrary to
the observations. 
We chose $V_\mathrm{in}=2000$\,\KMSEC for the lower limit of the infall velocity
because mergers with smaller infall velocities result in an integrated SZ effect 
($Y_{2500}$) that is too large to match the Bolocam measurements.
An infall velocity of $V_\mathrm{in} = 4000$\,\KMSEC already results in  
an unacceptably long time interval between the first and second core passages 
(longer than the age of the universe) for highly massive clusters.
Since such massive clusters have lower concentration parameters on average, 
we chose the lower limit of 5 and ran simulations within the range
$c_\mathrm{vir} \in [5,8]$. 
A full parameter search is not feasible with CPU clusters due to the 
large demand on computer resources.

In Figure~\ref{F:SIMUBAD}, we show mock \CHANDRA images 
based on our simulations between the second and third core passages
that do not provide the best match with the observations.
Most of the initial parameters were the same as in our best model (RP100V3.5), 
but we changed one or two parameters for each run to show how 
the changes impact the X-ray morphology (for the initial parameters, see
Table~\ref{T:TABLE1}).
The first  panel in the first row shows our simulation with an impact parameter 
changed to $P=200$\,kpc (RP200V3.5) at an epoch 1.1\,Gyr (8.7\,Gyr) 
after the second (first) core passage. 
The X-ray morphology shows only one bright peak. Thus, it does not 
match the observations although it has an integrated  
SZ amplitude of $Y_{2500} = 1.1 \times 10^{-10}$\,ster, which agrees
with the Bolocam constraint.

The second panel displays our simulation with the infall velocity
changed to $V_\mathrm{in}=3000$\,\KMSEC (RP100V3.0) at an epoch
0.68\,Gyr  (3.4\,Gyr) after the second (first) core passage. 
The X-ray surface brightness has two peaks with similar amplitudes 
(but no dominant peak) and no extension to the southern mass center.
Moreover, it also has a somewhat large integrated SZ amplitude,
$Y_{2500} = 1.4 \times 10^{-10}$\,ster.

In the third and fourth panels,
we show X-ray images based on our simulation with an infall velocity of  
$V_\mathrm{in}=3700$\,\KMSEC (RP100V3.7) at two different epochs,
0.32\,Gyr (11\,Gyr) and 1.4\,Gyr (12\,Gyr) after the second (first) core passage.
Again, the X-ray morphologies do not match with the observations:
the image in the third panel shows a single peak, while the fourth panel  
shows two peaks with similar amplitudes and no extension toward 
the southern mass peak. The integrated SZ amplitudes are 
$Y_{2500} = 2.6\times 10^{-10}$\,ster and $9.4 \times 10^{-11}$\,ster
at these epochs, and thus the first epoch can be excluded based
on its large integrated SZ amplitude as well.

The first and second panels in the second row of Figure~\ref{F:SIMUBAD}
display our simulations with concentration parameters $c_1 = c_2 = 6$
(RP100V3.5b) and $c_1 = c_2 = 8$ (RP100V3.5c), 
while the other initial parameters are the same as those of our best
simulation. 
These images are shown at epochs of 0.16\,Gyr (7.8\,Gyr) 
and 0.92\,Gyr (6.4\,Gyr) after the second (first) core passage.
The X-ray morphologies of these images do not provide a good
match with the observations, even though they satisfy all other
requirements including their integrated
SZ amplitude, $Y_{2500} = 1.0$ and $1.1\times 10^{-10}$\,ster, which
agree with the Bolocam constraint.

The third panel in the second row shows our simulations with 
an impact parameter of $P=300$\,kpc and an infall velocity of
$V_\mathrm{in}=3000$\,\KMSEC (RP300V3.0), at an epoch 
0.20\,Gyr (4.3\,Gyr) after the second (first) core passage. 
The X-ray morphology of this run shows two X-ray peaks, 
whereas the morphology does not provide a good match with the
observations.  
The fourth panel displays our simulations with an impact parameter of
$P=400$\,kpc, an infall velocity of $V_\mathrm{in}=3000$\,\KMSEC, and
concentration parameters of  $c_1 = 6$ and $c_2 = 7$ (RP400V3.0) at an
epoch 0.32\,Gyr (4.2\,Gyr) after the second (first) core passage. 
Again, we find two X-ray peaks, but the morphology does not match 
the observations.
These two runs can be excluded based on their large integrated SZ amplitudes, 
$Y_{2500} = 2.1$ and $2.4\times 10^{-10}$\,ster, which are more than
twice as large  as the Bolocam constraint, $Y_{2500}=0.9 \times
10^{-10}$\,ster.

On the basis of our simulations with fixed initial masses 
($M_1 = 1.7 \times 10^{15}M_\odot$ and $M_2 = 1.6 \times 10^{15}M_\odot$) 
and different impact parameters, infall velocities, and concentration parameters, 
we can provide a crude estimate on the uncertainties of our best model
parameters as
$P = 100_{-100}^{+100}\,$kpc,
$V_\mathrm{in} = 3500_{-500}^{+200}$\,\KMSEC, and 
$c_1 = 6\pm1$ and $c_2 = 6\pm1$. 
However, we should consider these initial conditions only as a guide
to set up the conditions for the second core passage, since 
our simulations have high fidelity only from around the second core
passage until the best epoch.
That is why we displayed the time elapsed from the
first core passage for our simulations only in parentheses. 
The reason for this is that, as usually done in controlled binary merger
simulations, the expansion of the universe and mass accretion from the
surrounding large-scale structure are ignored
(e.g., \citealt{RickerSarazin2001,RitchieThomas2002,PooleET2006,McCarthyET2007,ZuHone2011};
for a discussion, see \citealt{Molnar2016}).
This is a good approximation for a few Gyrs, but not for a significant
fraction of the age of the universe, such as 5--10\,Gyrs, 
which is the time scale of our simulations from the first to the second
core passage, due to the large masses and infall velocities we had to assume. 
On a time scale of a significant fraction of the age of the universe,
the expansion of the background and the mass accretion of clusters 
cannot be ignored. A study of the dynamics of A370 through its evolution 
would need a cosmological \NBODYHYDRO simulation, which would 
incorporate these effects naturally.
This could be done, for example, using constrained realizations of Gaussian 
random fields \citep{HoffmanRibak1991}, or one of its variants
(e.g., \citealt{RothET2016,ReyPontzen2018,SawalaET2020arXiv2003}).
However, this is out of the scope of our paper.

%
%
\begin{figure}[t]
\includegraphics[width=0.233\textwidth]{\FIGURES/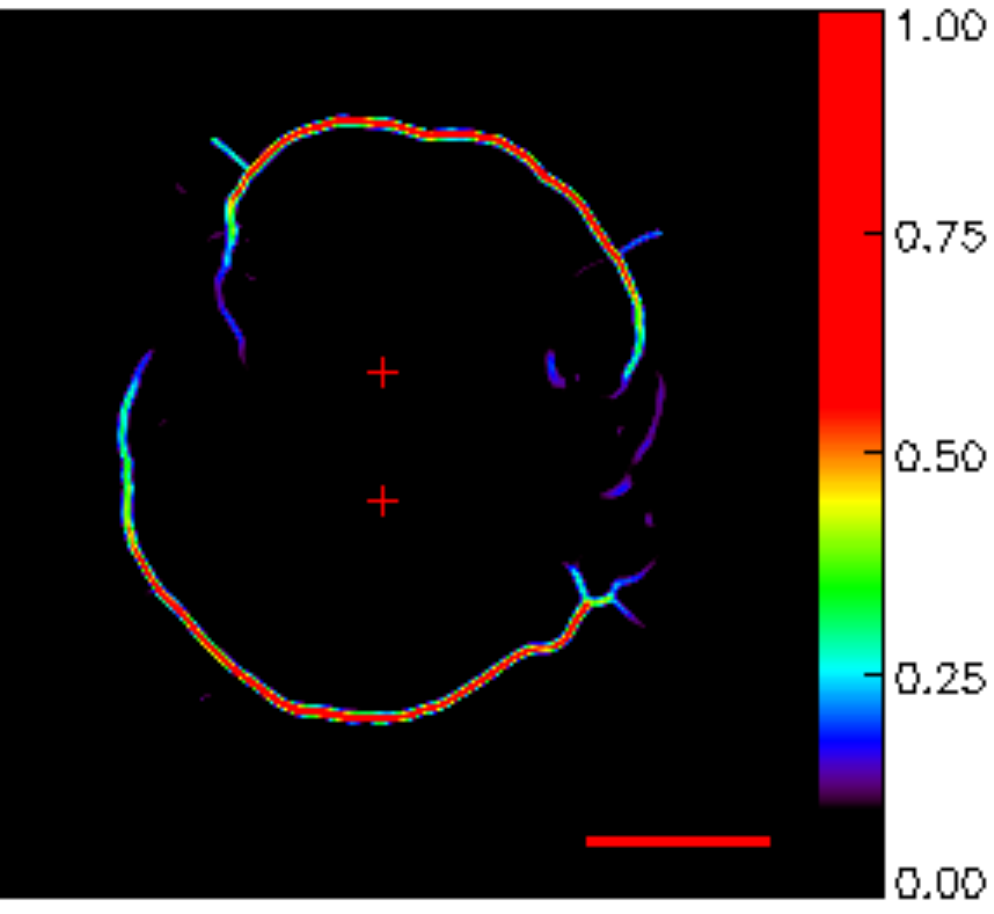}
\includegraphics[width=0.233\textwidth]{\FIGURES/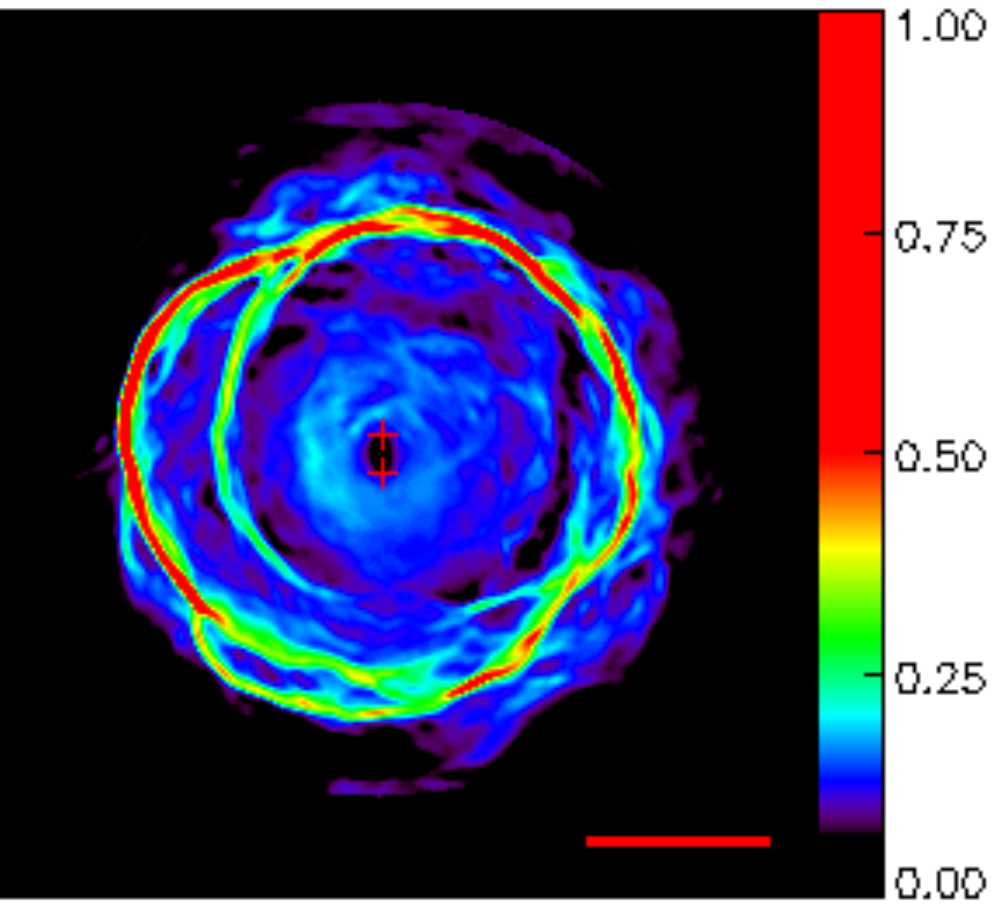}
\caption{
Images of shock positions (red curves) based on our best simulation run
(RP100V3.5) at the best epoch.
The two red crosses mark the peak locations of the mass surface density distribution.
The length of the horizontal red line is 1\,Mpc.
Left panel:
magnitude of the pressure gradient in a 2D cut through 
a plane containing the two cluster centers before rotation
out of the plane of the sky.
Right panel:
the magnitude of the pressure gradient of the projected pressure 
based on our best simulation with the best viewing angle
(Section~\ref{S:RESULTS}).
\label{F:SHOCKS}
}
\end{figure} 

In general, during each core passage, outgoing shockwaves are generated  
in the central region of merging galaxy clusters. 
We show in Figure~\ref{F:SHOCKS} the outgoing shocks generated after the second core passage
based on our best simulation (RP100V3.5) at the best epoch. 
We identified the shocks using the magnitude of the pressure gradient. 
The locations of large gradients in the pressure delineate the shock surfaces. 
In these images, the red curves show the location of the shocks.
The red crosses mark the dark-matter centers and 
the horizontal red bars indicate the physical scale of 1\,Mpc.

In the left panel of Figure~\ref{F:SHOCKS}, we show the magnitude of 
the pressure gradient in a two-dimensional (2D) cut through the centers of
the two components, before the rotation of the system out of the plane
of the sky. 
This image shows the physical (not projected) locations of the shocks.
The two shocks generated after the second core passage
can be clearly seen. The shocks are moving outward from the center.
Using this 2D pressure cut, we can derive the Mach numbers of these
shocks to obtain ${\cal M} \sim 3.2$.
At this epoch, the merging cluster is in an infalling phase,
where the two dark-matter components already turned over and 
are moving toward each other, approaching the third core passage.

In the right panel of Figure~\ref{F:SHOCKS},
we display the magnitude of the projected pressure gradient
using our best model, which is rotated out 
of the plane of the sky with a polar angle of $72.4^\circ$.
This panel shows where the shocks could be observed.
The two well-separated shock waves form entangled circles in projection.
We find that the closest shock surface to the cluster center is 
on the east side, about 890\,kpc away from the center. 
We note that, owing to projection effects,
the change in the projected pressure is not equal to the
physical pressure change due to the shock.
The Mach number derived from the pressure drop in this projection is 
${\cal M} \sim 1.2$, much less than the physical Mach number. 
We expect that observations of this outgoing shock would
measure a Mach number close to this value 
(for a discussion on how much bias is caused by projection effects
in the Mach numbers derived from X-ray observations,
see, e.g., \citealt{MolnarBroadhurst2017}).
The position of the closest shock in the east is in rough agreement 
with the position of the X-ray surface brightness edge we found by
reanalyzing the \CHANDRA observations of A370 
(at $\sim 690$\,kpc; Section~\ref{SS:XRData}; see Figure 5b).
Our simulations suggest that the X-ray surface brightness edge found in A370 
on the east from the cluster center is an outgoing shock generated after 
the second core passage.
According to our best model, the outgoing shock on the west of the
cluster center is located much farther out than the X-ray surface
brightness edge found by \cite{BotteonET2018MNRAS476}. 
Reanalyzing the \CHANDRA data, we found no significant X-ray edge on the
east from the cluster center, which is in agreement with our best model.

%
%
\begin{figure}[t]
\includegraphics[width=0.475\textwidth]{\FIGURES/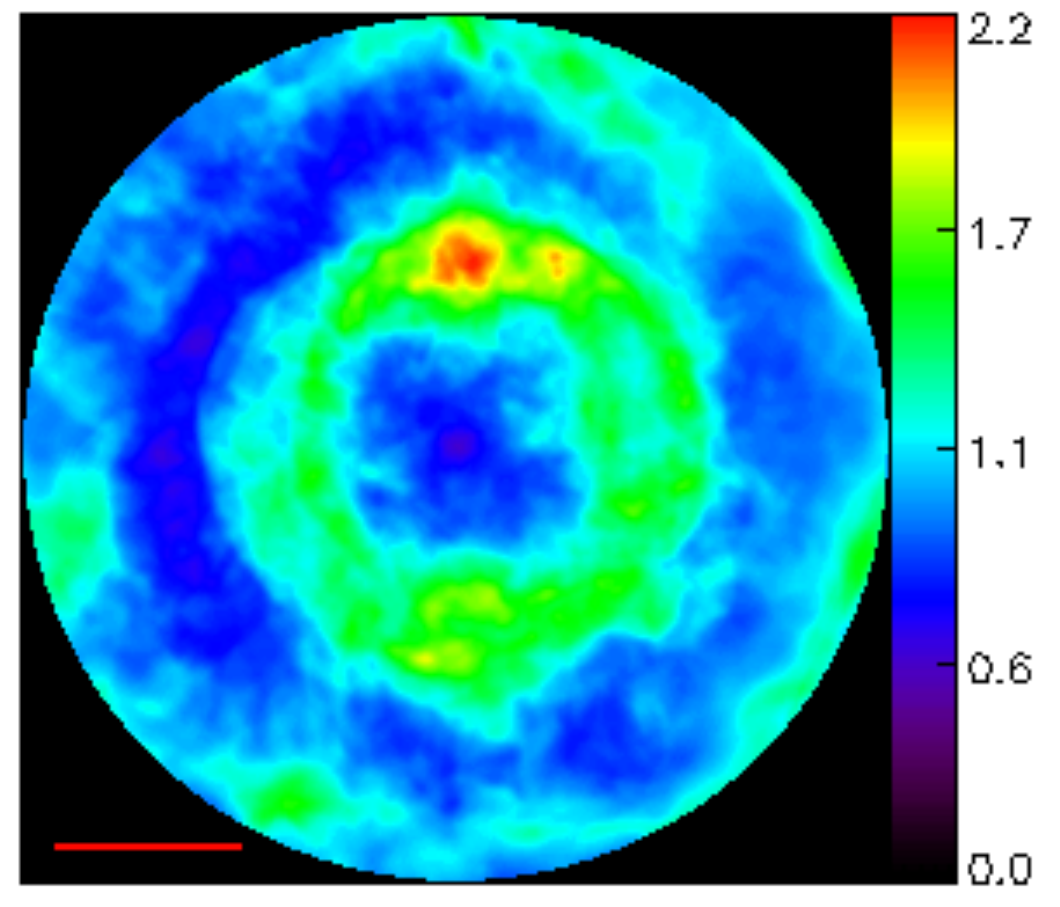}
\caption{
The gas mass surface density ratio, $\Sigma_\mathrm{best}/\Sigma_\mathrm{2CP}$ 
(best epoch over second core passage), based on our best model (RP100V3.5) 
at the best viewing angle within 2.5\,Mpc from the cluster center.
The red horizontal bar marks 1\,Mpc.
\label{F:BOLOCAM}
}
\end{figure} 

As a result of the outgoing shocks generated after the second core
passage, the gas density decreases in the central region. 
In a later phase, the gas falls back toward the center. 
We display in Figure~\ref{F:BOLOCAM} the ratio
$\Sigma_\mathrm{best}/\Sigma_\mathrm{2CP}$ 
between the gas mass surface density $\Sigma_\mathrm{best}$
at the best epoch (i.e., our best model) and that $\Sigma_\mathrm{2CP}$
at the epoch of the second core passage based on our best run
(RP100V3.5). These projected maps are obtained using the viewing angle
of our best model within 2.5\,Mpc from the cluster center. 
The red horizontal bar in the lower-left corner marks a scale of 1\,Mpc, 
which is about the size of the BOLOCAM extraction region
around the cluster center ($R \simlt R_{2500}$; central blue region). 
At the best epoch after the second core passage, 
the cluster passed the second turnover, while the gas is still moving
outward as a result of the shock, 
which is located about $1$--$2$\,Mpc from the cluster center 
(see Figure~\ref{F:SHOCKS}).
Accordingly, the gas is depleted in the central area where
$\Sigma_\mathrm{best}/\Sigma_\mathrm{2CP}\sim 0.7$   (dark blue region).  
This depletion of the gas from the central part of the system explains why
the X--ray emission and SZ amplitude of this cluster is well below those
suggested by cluster mass scaling relations (Section~\ref{S:Intro}).

\section{Summary}
\label{S:Summary}

We have used our existing library of binary merging cluster simulations
performed in our previous work and carried out self-consistent \NBODYHYDRO 
simulations with \FLASH to study the dynamical state of the massive 
merging cluster A370.
The cluster is a superlens system characterized by its large Einstein radius.
Our simulations were constrained by X-ray, SZ-effect, gravitational
lensing, and optical spectroscopic observations.
Specifically, we utilized the locations of the two mass peaks derived from 
strong gravitational lensing \citep{StraitET2018ApJ868}, 
the X-ray morphology (Section~\ref{SS:XRData}), 
the relative LOS velocity (Section~\ref{S:OpticalData}), and 
the amplitude of the integrated SZ effect
(Section~\ref{S:RadioData}; \citealt{CzakonET2015}).

We preformed \FLASH simulations by fixing the masses of the two 
cluster components based on weak- and strong-lensing measurements
\citep{Umetsu2011a}. We used different impact parameters, 
infall velocities, and concentration parameters to find the best match 
with the multi-wavelength observations.
We found that our best model with masses of 
$M_1=1.7 \times 10^{15} \,M_\odot$ and
$M_2=1.6 \times 10^{15} \,M_\odot$, 
an impact parameter of $P=100$\,kpc, an infall velocity of
$V_\mathrm{in}=3500$\,\KMSEC, and concentration parameters of $c_1 = 6$ and $c_2 = 6$
can explain the main features of the X-ray morphology and the mass
surface density
and simultaneously satisfy the constraints from
the observed relative LOS velocity and the integrated SZ amplitude.
Moreover, our best model reproduces the observed velocity dispersion of
cluster member galaxies, which supports the large total mass of A370
derived from weak lensing.
Our results strongly suggest that A370 is a post major merger
observed 0.57\,Gyr after the second core passage, just before the third
core passage.

It is intriguing to note that, similar to the case of A370,
Cl0024$+$1654 is another superlens cluster at $z=0.395$ that is very
faint in X-ray and SZ signals \citep{Umetsu2011a,Umetsu2011b}. 
A careful interpretation of X-ray, lensing, and dynamial
data based on \FLASH simulations suggests that Cl0024$+$1654 is also
a post major merger occurring along the line of sight, viewed
approximately 2--3\,Gyr after impact before the gas has recovered
\citep{Umetsu2010}.

Our simulations represent the first attempt to model one of 
the most massive dynamically active merging galaxy clusters, A370, 
using dedicated self-consistent \NBODYHYDRO simulations. 
In order to improve our dynamical model for A370, deeper X-ray observations 
would be necessary to verify the positions of the mean and secondary peaks,
obtain a spatially resolved temperature distribution of the intracluster gas,
and verify the shock feature in the east from the cluster center. 
Moreover, a detailed optical spectroscopic study would be needed to
precisely determine the relative LOS velocity of the two main merging
components. 
With these proposed new observations a more extensive parameter 
search would be worth while to pursue, which is more likely feasible
using GPU clusters.

\acknowledgements
We thank the referee for a thorough reading of our manuscript, and 
for comments and suggestions, which made the presentation of our
results clearer. We are grateful to J. Richard and D.~J., Lagattuta
for providing us with the electronic table of the cluster redshifts from
their latest survey of A370 and helping with analyzing the data.
The code \FLASH\ used in this work was in part developed by the
DOE-supported ASC/Alliance Center for Astrophysical Thermonuclear
Flashes at the University of Chicago.  
We carried out our simulations using the high performance computer facility 
at the Academia Sinica Institute of Astronomy and Astrophysics.
This work was supported in part by the Ministry of Science and Technology 
of Taiwan (grant MOST 106-2628-M-001-003-MY3) and by Academia Sinica 
(grant AS-IA-107-M01).

%
%
\bibliographystyle{apj}


\end{document}